\documentstyle[twocolumn,aps,amsfonts,floats,epsf]{revtex}

\begin{document}


\twocolumn[\hsize\textwidth\columnwidth\hsize\csname %
@twocolumnfalse\endcsname

\draft

\title{
  Material-specific gap function in the high-temperature
  superconductors}

\author{
  B.~E.~C.~Koltenbah and Robert Joynt$^\dagger$}


\address{
  Department of Physics and Applied Superconductivity Center,
  University of Wisconsin-Madison\\
  1150 University Avenue,
  Madison, Wisconsin  53706\\
  $^\dagger$ and
  Materials Physics Laboratory,
  Helsinki University of Technology,
  FIN-02150 Espoo, Finland}

\date{May 28, 1996}

\maketitle


\begin{abstract}
We present theoretical arguments and experimental support  for the
idea that high-$T_c$ superconductivity can occur with $s$-wave,
$d$-wave, or mixed-wave  pairing in the context of a magnetic
mechanism.  The size and shape of the gap 
is different for different materials. The
theoretical arguments are based on the $t$-$J$ model as derived from
the Hubbard model so that it necessarily includes three-site terms. 
We argue that this should be the basic minimal model for high-$T_c$
systems. We analyze this model starting with the dilute limit which
can be solved exactly, passing then to the Cooper problem which is
numerically tractable, then ending with a mean field approach. It is
found that the relative stability of $s$-wave and $d$-wave depends on
the size and the shape of the Fermi  surface.  We identify three
striking trends.  First, materials with large next-nearest-neighbor
hopping  (such as YBa$_2$Cu$_3$O$_{7-x}$) are nearly pure $d$-wave,
whereas nearest-neighbor materials (such as La$_{2-x}$Sr$_x$CuO$_4$)
tend to be more $s$-wave-like. Second, low hole doping materials tend
to be pure $d$-wave, but high hole doping leads to $s$-wave.  Finally, 
the optimum hole doping level increases as the next-nearest-neighbor
hopping increases.  We examine the experimental evidence and find
support for this idea that gap function in the high-temperature
superconductors is material-specific.
\end{abstract}

\pacs{PACS numbers: 74.20.De, 74.20.Mn, 74.72.-h}

]


\section{Introduction}
\label{sec:intro}

In the years following the discovery of the high-$T_c$ materials,
theoretical debate often centered on the  anomalous normal state
properties, these being seen as the  key to understanding the
underlying physics.  The superconducting state, by contrast, was
generally  thought to be conventional except for its coupling
strength. More recently, however, it has seemed more  reasonable to
regard  the superconducting state, especially the gap symmetry, as
holding the key to theoretical understanding.  Recent experimental
progress gives hope  that the gap symmetry can be unraveled.  Once
this is done, one expects that very strong constraints can be placed
on the microscopic model. This will certainly be true if, as we shall
contend in this paper, different high-$T_c$ systems have different gap
functions.

The early consensus that the high-$T_c$ superconductors  are $s$-wave
has been replaced by the view that they are most likely $d$-wave. 
Support for this latter position has come primarily from three
experimental sources: penetration depth measurements
\cite{hardy,maryland}, photoemission measurements  \cite{ma,shen}, and
Josephson interference measurements \cite{illinois,ibm,eth}.  These
results very much strengthened the idea that the mechanism is
magnetic.  Even before the  discovery of high-$T_c$ superconductivity
it  was clear that magnetic interactions based on antiferromagnetic
correlations would quite naturally give rise to a tendency toward
higher-wave pairing \cite{varma,emery}, and studies of the $t$-$J$
model \cite{ruckenstein,gros} and the Hubbard model
\cite{bickers} around the time of  the discovery  confirmed this for the
CuO$_2$ square lattice . A more complete
theory, though one which requires some phenomenological input, has
been constructed on the hypothesis that antiferromagnetic spin
fluctuations act very much like the phonons in low-$T_c$ materials and
cause the pairing \cite{ueda,pines}.  This theory, which we shall call
the spin-fluctuation model, requires the presence of very strong spin
correlations which lead to a high critical temperature and to
$d$-wave pairing.  
 
If it is accepted that a magnetic mechanism is responsible  for
high-$T_c$, there still remain many outstanding issues. The
spin-fluctuation model generally relies on a pairing interaction which
is very well localized in {\it momentum} space.  The interaction is
proportional to the magnetic susceptibility, which is taken to be very
strongly peaked near the $k_x = \pm \pi/a$ and $k_y = \pm \pi/a$
points.  The spin-fluctuation model leads unambiguously to $d$-wave
superconductivity and not to $s$-wave.  This also appears to be the case in 
the spin bag model, although the range of the interaction is
shorter in this case, of order $\hbar v_F / \Delta_{\text{SDW}}$,
where $v_F$ is the Fermi velocity and $\Delta_{\text{SDW}}$ is the 
charge density wave gap of the ordered phase \cite{bobs}.
On the other hand, theories
based on the $t$-$J$ model rely on a spin-spin interaction which is
rather local in {\it real} space.  We shall call this the
spin-interaction model. If the gap equation is solved in this model,
substantial regions of both $s$-wave and $d$-wave pairing are found
\cite{ruckenstein}. In earlier variational Monte Carlo (VMC)
calculations on this model, the dominant instability of the normal
state was toward $d$-wave pairing, but in some parameter  regimes
there was also the possibility of admixture of $s$-wave symmetry into
the ground state \cite{koltenbah}. Thus, although  the
spin-interaction model clearly falls into the category of magnetic
mechanisms, it is distinct from the spin-fluctuation model in that it
predicts that the gap symmetry is not necessarily 
always pure $d$-wave.  Rather,
$s$-wave and $d$-wave are competing instabilities. 
This latter direction of inquiry has been briefly summarized by 
M\"{u}ller, who notes that the
possible coexistence of $s$- and $d$-wave can describe recent
experimental results \cite{muller}.  Our Sec.~\ref{sec:survey}
is an expansion of this theme.  

The comparison of
the spin-fluctuation model and the spin-interaction model 
leads to the conclusion that all magnetic mechanisms
are {\it not} alike.  Even if it is conceded that the basic mechanism
for high-$T_c$ superconductivity is magnetic, there is still work to
be done.  This paper is written to amplify and sharpen this point. 

We present a brief discussion of recent theoretical 
work which is directly relevant to the issue
material-specific gap functions (Sec.~\ref{sec:theory}) arising from
the magnetism mechanism. The
extended $t$-$J$ model is developed from the
overlapping copper and oxygen orbitals in the CuO$_2$ planes
(Sec.~\ref{sec:model}). We point out physical reasons to support the
idea of $s$- and $d$-wave as competing instabilities based upon the
solution of the $t$-$J$ model in the dilute limit
(Sec.~\ref{sec:dilute}). It is noted that this solution in fact
corresponds to a physical limit of the gap equation in which the
interactions do not possess a frequency cutoff. We then extend our
arguments by solving the Cooper problem with the $t$-$J$ model
(Sec.~\ref{sec:cooper}). We follow with a presentation of a mean field
approach to the $t$-$J$ model which does indeed yield mixed $s$- and
$d$-wave pairing under certain circumstances
(Sec.~\ref{sec:meanfield}).  Finally, we
survey existing experiments and find evidence of the trends identified
in the analysis of our model (Sec.~\ref{sec:survey}). The approach of
the experimental survey is to examine different materials one by one
in order to see if they all have the same gap symmetry.   Our
conclusion is that the details of the gap function are material-specific, and that
this lends support for using the $t$-$J$ model, extended in an
appropriate fashion, as a basic, yet flexible description of
high-temperature superconductivity in the cuprate-layered materials.


\section{Theoretical Background}
\label{sec:theory}

Admirably complete and comprehensive surveys of theoretical work
on the magnetic mechanism and the question of d-wave pairing
has recently been carried out by Scalapino \cite{doug}  
(see particularly Appendix A, and references therein),
Dagotto \cite{elrmp},  and von der Linden \cite{vdl}, who stress the
numerical work.  
They conclude that numerical 
evidence from quantum Monte Carlo, variational Monte
Carlo, and exact diagonalization on small systems
favors d-wave pairing. 
Certain diagrammatic studies confirm this.
These references also point out that the 
evidence is far from conclusive.  
For example, quantum Monte Carlo searches for
a finite superfluid density in the 
two-dimensional Hubbard model have not been successful.
Thus, studies of d-wave
superconductivity in magnetic models
are highly suggestive, but we are not assured
that superconductivity exists in these microscopically justified
models.

Our approach in this paper is to sidestep this thorny
issue.  Instead of investigating the microscopic models with the
most sophisticated tools available to ask whether superconductivity
is present, we shall use simple mean-field-type
methods which are 
expected to give superconductivity.  The idea is to
apply these methods to 
rather more complicated models, intended to more closely
mimic the actual systems.  If we can identify trends as a function of the 
(fairly numerous) parameters in the models, these may then be compared to
experiment in a detailed way.  Accordingly, we do not review
those studies which have been
carried out on the nearest-neighbor Hubbard model
or its strong-coupling equivalent, the nearest-neighbor
$t$-$J$ model, the aim of which has generally been to
search for pure d-wave superconductivity.  
Our interest is in the question of whether the 
gap structure may be material-specific. 
Specifically, we wish to address the issue of whether the
shape and size of the gap function may depend on details
of the band structure and the doping level.  We therefore
review here those papers which treat more complex models 
or more complicated gap functions.
                                                                   
The finding of a more complex gap structure
in the $t$-$J$ model originates from the very beginning of this field of 
study.  Ruckenstein {\it et al.}, were among the first to apply this 
model for use in calculating some of physical parameters of the 
high-temperature superconductors \cite{ruckenstein}.  They 
analyzed the model with a mean field treatment (similar to 
Sec.~\ref{sec:meanfield}), and found a parameter range
where a mixture of $s$-wave and $d$-wave 
was the most stable ground state.

Gros, {\it et al.}, used a
variational Monte Carlo method by which the ground state of
the $t$-$J$ model was studied 
without the need for further approximations such as mean field \cite{gros}.  
This method has the advantage of explicitly 
handling the requirement of no double-occupancy.  Furthermore, this method 
is a test for evaluating candidate 
wavefunctions and comparing their ground state energies with one another 
through extensive parameter surveys.

Li and the present authors utilized this VMC method in 
studying the $t$-$J$ model along with the three-site terms 
to suggest the possibility of $sd$-mixing in the cuprates \cite{koltenbah}.
In this paper, we presented VMC calculations which compared ground state 
energies of $d$-wave, extended $s$-wave and various mixtures of the 
two.  The results showed, as is repeated often throughout this paper, 
that $d$-wave was the ground state near half-filling, whereas an 
admixture of $s$ and $d$ won out at higher hole doping.  
This VMC work, done solely on 
tetragonal lattices, found little difference in energy
among the various kinds of $sd$-mixing, namely $s$+$d$, 
$s$+$id$, or various phases in between.  Rather, all such 
mixtures were deemed degenerate within the statistical error of the 
calculations, as can clearly be seen in Fig.~6(a) of that work.
We proposed at the time that the introduction of 
anisotropy would lift this degeneracy, and indeed recent calculations 
indicate (with allowance for further necessary study) that in orthorhombic
lattices, the VMC calculations favor an 
$s$+$d$-wave state \cite{unpub}.
These are preliminary results of a nonsystematic study of 
the $t$-$J$ model phase diagram, the details of which are not clear at 
this time and demand further analysis before more definite claims can be 
made.

The recent work of O'Donovan and Carbotte yields $s$- 
and $d$-wave mixing \cite{carbotte2}.  Using a tight-binding model with orthorhombic 
distortion, they solved the zero-temperature BCS gap equation with a nearest-neighbor 
interaction, and found, as we do throughout this work, that only 
$s_{x^2+y^2}$ and $d_{x^2-y^2}$ components are found.  Indeed, nearer to 
half-filling, they saw a predominant $d$-wave with small $s$-wave 
admixture, whereas at smaller density $n$, the $s$-wave component increased 
until it was the predominant phase with small $d$-wave admixture.  The 
relative phases between the two was $0$ (a $d$+$s$-wave phase) shifting 
rather quickly as $n$ decreased further to $\pi$ ($s$-$d$-wave phase) with the increase of the 
$s$-wave component.  This cross-over of predominant $d$- to $s$-wave 
is in rough agreement with the trends noted in \cite{koltenbah}
and also below.

B\'{e}al-Monod and Maki also solved the ze\-ro-tem\-per\-a\-ture gap equation 
using a fermion-fermion interaction, where only states on the Fermi surface
were considered in their analysis \cite{bealmonod}.  They found that 
even a small anisotropy enhanced the maximum gap as well as the 
transition temperature.  In the tetragonal case, they found pure $d$-wave 
as the solution.

Varelogiannis, in study of electron-phonon interactions in the 2D 
superconductors, has found that either $s$- or $d$-wave results from 
consideration of the Coulomb pseudopotential as well as doping 
\cite{varelog}.  That some electron-phonon interaction investigations 
are yielding similar qualitative results as spin-interaction models
is interesting.

The work most similar in spirit to the present is
that of Dagotto and collaborators.  They have also 
stressed that spin-interactions which are short-range in 
real space may lead to a rich superconducting
phase diagram with a region of stability for $s$-wave \cite{elbio}.
These references also make the important point that 
details of the band structure, particularly peaks in the
density of states, may have an important influence 
on superconductivity.  The band structure dependence is 
also one of the important aspects of the results 
below; we find other, more local, aspects of the
band structure to be important as well.  

Scalapino's survey indicates that most theoretical work on magnetic mechanisms
has concentrated on the issue
of ``$s$-wave versus $d$-wave": do the simplest models 
give rise to $d$-wave superconductivity?   The whole range of
theoretical tools available has been applied to this problem.
If they do, and if the experiments demonstrate that the high-$T_c$
materials are $d$-wave, then, taking everything 
together, we have evidence for
a magnetic mechanism.  Taking our cue from the papers
which show more complex phase diagrams arising
from magnetic interactions, we wish to add an additional
element to this debate.  Is it possible to identify 
{\it trends} in the theoretical results on the size and shape
of the gap function which can be compared with experiment?
Our strategy will be to look at the simplest magnetic model,
the $t$-$J$ model, but adding enough flexibility to the model
that we can hope to make material-specific predictions.
The model then becomes sufficiently complicated that
we are limited to the simplest mean-field-type methods.        
These methods do have the advantages of simplicity 
and physical transparency.



\begin{figure}[ht]
\begin{center}
\leavevmode
\epsffile{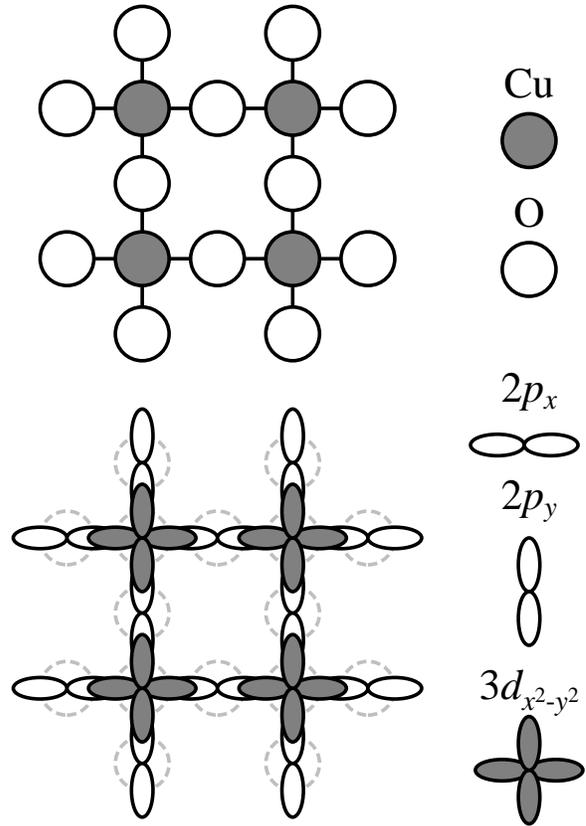}
\caption[]{
Diagram of the CuO planes showing four unit cells and the overlay 
of the Cu $3d_{x^2-y^2}$ and O $2p_x$ and $2p_y$ orbitals.  Note 
that a more realistic picture would show the actual hybrid orbitals
of the high-$T_{c}$ CuO planes.}
\label{fig:CuO4}
\end{center}
\end{figure}

\section{The Model}
\label{sec:model}

In high-$T_c$ systems, conduction takes place in the copper-oxygen
planes. The unit cell consists of one copper and two oxygen atoms as
seen in Fig.~\ref{fig:CuO4}. The chemistry of these atoms suggest, and
band  structure calculations confirm, that there is only one orbital
on each site that plays a role in this conduction.  These are the
oxygen $2p_x$ or $2p_y$ orbitals ($x$ or $y$ depending on which one
points to the neighboring copper atoms) and the  copper $3d_{x^2-y^2}$
orbital.  The six states must accommodate five electrons in, for
example, La$_2$CuO$_4$.  We shall call this the half-filled band case,
for reasons which will become clear later. In La$_{2-x}$Sr$_x$CuO$_4$,
they must  accommodate $5 - x$ electrons, as Sr contributes 1 less
electron per unit  cell to the CuO planes.  The noninteracting model
is easily solved.

Let us take creation operators $d^\dagger_{\bf r}$, $p^\dagger_{{\bf
r}1}$, and $p^\dagger_{{\bf r}2}$.  Then the Hamiltonian is
\begin{eqnarray}
  {\cal H} & = &
  \sum_{\bf r} \varepsilon_d d^\dagger_{\bf r} d^{}_{\bf r} +
  \sum_{\bf r} \varepsilon_p
    (p^\dagger_{{\bf r}1} p^{}_{{\bf r}1} +
     p^\dagger_{{\bf r}2} p^{}_{{\bf r}2}) \nonumber \\
  & &
  - t_{pd \sigma} \sum_{\bf r}
    \left[ (d^\dagger_{\bf r} p^{}_{{\bf r}1} +
            p^\dagger_{{\bf r}1} d^{}_{\bf r}) -
           (d^\dagger_{\bf r} p^{}_{{\bf r}2} + 
            p^\dagger_{{\bf r}2} d^{}_{\bf r}) \right] \nonumber \\
  & &
  + t_{pd \sigma} \sum_{\bf r} 
    \left[ (d^\dagger_{{\bf r} + a\hat{x}} p^{}_{{\bf r}1} + 
            p^\dagger_{{\bf r}1} d^{}_{{\bf r} + a\hat{x}}) \right.
            \nonumber \\
  & & ~~~~~~~~~~~
    \left.-(d^\dagger_{{\bf r} + a\hat{y}} p^{}_{{\bf r}2} + 
            p^\dagger_{{\bf r}2} d^{}_{{\bf r} + a\hat{y}}) \right]. 
\end{eqnarray}
Here, $\varepsilon_d$ and  $\varepsilon_p$ are the atomic energies of
the two orbitals and
\begin{equation}
  t_{pd \sigma} \equiv - \int d^3r ~ \psi_d({\bf r}) ~ 
  \Delta{\cal H} ~ \psi_{py}({\bf r} - a\hat{y})  
\end{equation} 
is the overlap integral of the $ 3d_{x^2-y^2} $ orbital and the $
2p_y$ orbital in the unit cell.  $ \Delta {\cal H} $ is the difference
of the  atomic and crystal Hamiltonians. All the other overlaps in our
restricted basis set are related to this one  by symmetry.  The
relative signs in the Hamiltonian are important, and are determined by
the character of the atomic wavefunctions. The above tight-binding
Hamiltonian assumes that the orbitals are orthogonal to each other.

The Bloch wavefunction is now defined by
\begin{eqnarray}
  \psi_{\bf k} & = &
  \frac{1}{\sqrt{N_c}} \sum_{\bf r}
    e^{i {\bf k} \cdot {\bf r}} \nonumber \\
  & & \times
    \left[ u_1 ~ p^\dagger_{{\bf r}1}({\bf k})
         + u_2 ~ p^\dagger_{{\bf r}2}({\bf k}) 
         + u_3 ~ d^\dagger_{\bf r}({\bf k}) \right]
    \left|0\right>,
\end{eqnarray}
with the $u_i({\bf k})$ still to be determined. The matrix elements of
${\cal H}$ are
\begin{equation}
  \left< \psi^*_{\bf k} | {\cal H} | \psi_{{\bf k}'} \right> =
  \delta_{{\bf k}, {\bf k}'} \sum_{i,j} ~
  h_{ij}({\bf k}) ~ u^*_i({\bf k}) ~ u_j({\bf k}),
\end{equation} 
with
\begin{eqnarray}
  & & h_{ij}({\bf k}) = \\
  & &
  \left( \begin{array}{ccc}
    \varepsilon_p & 0 & t_{pd \sigma} (\text{-}1 + e^{ik_ya}) \\
    0 & \varepsilon_p & t_{pd \sigma} (1 - e^{ik_xa}) \\
    t_{pd \sigma}(\text{-}1 + e^{-ik_ya}) & 
    t_{pd \sigma}(1 - e^{-ik_xa}) &  \varepsilon_d 
  \end{array} \right) \nonumber
\end{eqnarray}
as may be verified by a simple calculation. The secular equation to
determine the eigenvalues 
$\varepsilon({\bf k})$ at a given 
wavevector is
\begin{equation}
  \left| h_{ij}({\bf k}) - \varepsilon({\bf k}) \right| = 0,
\end{equation}
which takes the form
\begin{eqnarray}
  & & (\varepsilon_p - \varepsilon({\bf k})) 
  \left[ (\varepsilon_p - \varepsilon({\bf k}))
         (\varepsilon_d - \varepsilon({\bf k}))
       - 4t_{pd \sigma}^2 \sin^2{\left( \frac{k_x a}{2} \right)}
  \right. \nonumber \\
  & & ~~~~~~~~~~~~~~
  \left. 
       - 4t_{pd \sigma}^2 \sin^2{\left( \frac{k_x a}{2} \right)}
  \right] = 0.
\end{eqnarray}
There are three solutions to this equation, which we shall call
$\varepsilon_n({\bf k})$, $\varepsilon_+({\bf k})$, and
$\varepsilon_-({\bf k})$.  They are
\begin{eqnarray}
  \varepsilon_n({\bf k}) & = & \varepsilon_p,
  \label{eq:disp1} \\
  \varepsilon_{\pm}({\bf k}) & = &
    \frac{\varepsilon_p + \varepsilon_d}{2} \pm \frac{1}{2}
  \left[(\varepsilon_d - \varepsilon_p)^2 
    + 16 t_{pd \sigma }^2 \sin^2{\left( \frac{k_x a}{2} \right)}
    \right. \nonumber \\
  & & ~~~~~~~~~~~~~~~~ \left.
      + 16 t_{pd \sigma }^2 \sin^2{\left( \frac{k_x a}{2} \right)}
    \right]^{1/2}. 
  \label{eq:disp2}
\end{eqnarray}
We see that the band corresponding to $\varepsilon_n({\bf k})$ is
completely flat.  It has no amplitude on the  Cu-atoms and can be
thought of as a nonbonding orbital. $\varepsilon_-({\bf k})$ has
dispersion which reaches downward from the atomic energies and is a
bonding orbital.   $\varepsilon_+({\bf k})$ has dispersion which
reaches upward from the atomic energies and is an antibonding orbital.  
It is generally found that $\varepsilon_p \approx \varepsilon_d $.
This leads to the very strong hybridization between  copper and oxygen
which is characteristic of the high-$T_c$ materials and distinguishes
them, chemically, from  most other transition-metal oxides.  The two
energies  lie about $3.2$ eV  below the Fermi energy.   (We shall not
attempt to assign precise values to these bare parameters, since the
observable parameters  are those of the effective model to be derived
below. They are the important ones, and they are best taken from
experiment.) With $t_{pd \sigma }  \approx 1.80$ eV, only the
antibonding band crosses the Fermi energy. At the Fermi energy, the
wavefunctions have substantial weight at both copper and oxygen sites.
 
With interaction, we must add a term
\begin{equation}
  {\cal H}_{\text{int}} =
    U_d \sum_{\bf r} n_{d{\bf r}\uparrow} n_{d{\bf r}\downarrow}
  + U_p \sum_{{\bf r}i} n_{p{\bf r}i\uparrow} n_{p{\bf r}i\downarrow}.
\end{equation}
Here the sum runs over all atoms in the plane, and the interaction is
approximated to be local; only electrons on the same atom interact. 
Theories of high-temperature superconductivity based on charge
fluctuation-mediated attractions go beyond this approximation by
including longer-range terms, particularly interactions between
electrons on  neighboring copper and oxygen atoms.  These interactions
are not so important for the magnetic mechanisms which are the subject
of this paper.
  
Since the parameter $U_d$ is large, ($\sim 10$ eV) we may expect  a
considerable revision of the energy levels of the noninteracting
problem to take place.  In the ground state of the  half-filled band
case, the copper orbital will have essentially only one electron,
since the addition of a second would cost this very large  amount of
energy.  This leads to an insulating state because any flow of charge
would require a real change of the occupations from their base values
of one per copper and two per oxygen. Such a state is in fact observed
in    La$_2$CuO$_4$.  It is also a magnetic state, in the sense that
every copper atom has spin one-half.  (The atomic configuration is
$3d^9$.)  

$U_p$ is smaller than $U_d$, but  large enough that holes added to the
half-filled band state go  predominantly to the oxygen sites.  Each
such site has then also a spin one-half.  It was shown by Zhang and
Rice \cite{zr} that these oxygen spins pair with the copper spins to
form a spin singlet.  This is the entity which moves from unit cell to
unit cell.  It produces metallic conduction, as it has a charge of
$-1$ relative to the rest of the lattice. It also has a spin of zero,
while the other cells have spin one-half because of their unpaired
copper spins. The Hamiltonian for these particles is
\begin{eqnarray}
  {\cal H}_{\text{ZR}} & = &
    -t \sum_{{\bf r},\bbox{\delta},\sigma} 
    c^\dagger_{{\bf r},\sigma} c^{}_{{\bf r}+\bbox{\delta},\sigma}
    -t' \sum_{{\bf r},\bbox{\gamma},\sigma} 
    c^\dagger_{{\bf r},\sigma} c^{}_{{\bf r}+\bbox{\gamma},\sigma}
    \nonumber \\
  & &
    + U \sum_{\bf r} n_{{\bf r} \uparrow} n_{{\bf r} \downarrow}.
\end{eqnarray}
We refer the reader to Ref.\ \cite{zr} for the details of the
derivation.   Here $c_{{\bf r} \sigma}$ creates a hole and ${\bf r}$
refers to the position of a unit cell, not an atom, and
$\bbox{\delta}$ and $\bbox{\gamma}$ are nearest-neighbor and
next-nearest-neighbor vectors, respectively, of the square lattice of
unit cells. This Hamiltonian is known as the single-band Hubbard
Hamiltonian. It has received a great deal of attention in connection
with the high-$T_c$ problem.  Our aim here is not to review this work,
but to do the simplest calculations which are relevant to
superconductivity in a certain limit of the model, namely the limit $U
>> t, U >> t'$. We shall assume that $t$ and $t'$ are of the same
order of magnitude.

The appropriate method in this limit is perturbative, in that the
interaction term ${\cal H}_{\text{int}}$ is taken as the unperturbed
Hamiltonian.  We have that  
\begin{equation}
  {\cal H}_{\text{int}} = U \sum_{\bf r}
  n_{{\bf r}\uparrow }n_{{\bf r}\downarrow } ,
  \label{eq:hubbU}
\end{equation}
and we wish to solve
\begin{equation}
  {\cal H}_{\text{int}} \Psi = E \Psi,
\end{equation}
to get started.  
Since ${\cal H}_{\text{int}}$ simply counts the number of
doubly-occupied sites, however, the basis of wavefunctions in
configuration space (where the occupation of each site is specified)
is already diagonal. The eigenvalues are just $0$, $U$, $2U$, and so
on, corresponding to $0$, $1$, $2$, etc., doubly-occupied sites.  Each
eigenvalue is  very highly degenerate.  Since our interest is in the
low energy states, let us concentrate on the subspace with no
doubly-occupied sites, with an unperturbed energy of zero.

Now turn on the perturbation, which is
\begin{equation}
  \sum_{{\bf k}s} \varepsilon_{\bf k} n_{{\bf k}s}. 
\end{equation}
The effect of this operator on a state  in configuration space is to 
move one electron from one site to a nearby site. In acting on a state
with no doubly-occupied sites, it may, for example, move an electron
to a site which is already singly-occupied, thereby increasing the
number of doubly-occupied sites by one. The kinetic energy operator
has the effect of mixing the different subspaces.  We shall take this
into account in the following way.  If an arbitrary wavefunction
$\Psi$ is written in configuration space, then it may be decomposed as
\begin{equation}
  \Psi = \Psi_0 + \Psi_1 + \Psi_2 + \ldots,
\end{equation}
where $\Psi_0$ has no doubly-occupied sites,
$\Psi_1$ has one doubly-occupied site,
and so on.

The Schr\"odinger equation may then be written as
\begin{equation}
  \left( \begin{array}{cc}
    \begin{array}{ccc}
      {\cal H}_{00} & {\cal H}_{01} & 0 \\
      {\cal H}_{10} & {\cal H}_{11} & {\cal H}_{12} \\
      0             & {\cal H}_{21} & {\cal H}_{22}
    \end{array} & \ldots \\
    \vdots & \ddots
  \end{array} \right)
  \left( \begin{array}{c}
    \Psi_0 \\
    \Psi_1 \\
    \Psi_2 \\
    \vdots 
  \end{array} \right) = E
  \left( \begin{array}{c}
    \Psi_0 \\
    \Psi_1 \\
    \Psi_2 \\
    \vdots
  \end{array} \right).
\end{equation}
Here the $\Psi$ are as defined above, and ${\cal H}_{nn'}$ is the part
of the Hamiltonian which acts only on a state with $n'$
doubly-occupied sites and produces a state with exactly $n$
doubly-occupied sites.  This Schr\"odinger equation is still
completely general.  If $U$ is large enough, and we are interested
only in the low-lying states,  we may neglect $\Psi_2$, and all states
with more than two doubly-occupied sites.  Then the first two
equations are
\begin{eqnarray}
  {\cal H}_{00} \Psi_0 + {\cal H}_{01} \Psi_1 & = & E \Psi_0, \\
  {\cal H}_{10} \Psi_0 + {\cal H}_{11} \Psi_1 & = & E \Psi_1.
\end{eqnarray} 

Substituting the second equation into the first, we find
\begin{equation}
  {\cal H}_{00} \Psi_0 + {\cal H}_{01} (E- {\cal H}_{11})^{-1}
  {\cal H}_{10} \Psi_0  =  E \Psi_0.
\end{equation}
The second term on the left-hand side has the following structure. 
${\cal H}_{10} \Psi_0 $ is a state with exactly one doubly-occupied
site.   Its energy is therefore $U$, with corrections of order $t$ or
$t'$.  ${\cal H}_{11}$ acting on this state gives $U$ to the
approximation in which we are interested, and we may write the entire
equation as 
\begin{equation}
  {\cal H}_{00} \Psi_0 - \frac{1}{U} {\cal H}_{01}
  {\cal H}_{10} \Psi_0  =  E \Psi_0.
\end{equation}
We have then a new Schr\"odinger equation to solve, which may be
written as
\begin{equation}
  {\cal H}_{\text{eff}} \Psi_0  =  E \Psi_0,
\end{equation}
with
\begin{equation}
  {\cal H}_{\text{eff}} = {\cal H}_{00} - \frac{1}{U} {\cal H}_{01}
  {\cal H}_{10}.
\end{equation}

In this equation, the second term represents virtual processes in
which a doubly-occupied site is first created, then destroyed.  In
second order perturbation theory (order $t^2$), these processes affect
the energies of the lowest energy states and break the degeneracy of
the lowest level.  Note that ${\cal H}_{00}$ also has kinetic energy
terms if the band is less than half filled, as it is possible then to
move electrons from one singly-occupied site to another, creating no 
doubly-occupied sites.  Explicitly, the various operators have the
form
\begin{eqnarray}
  {\cal H}_{10} & = &
  - t \sum_{{\bf r},\bbox{\delta},\sigma}
    n^{}_{{\bf r},-\sigma}
    c^\dagger_{{\bf r},\sigma} c^{}_{{\bf r}+\bbox{\delta},\sigma}
    (1 - n^{}_{{\bf r}+\bbox{\delta},-\sigma}) \nonumber \\
  & &
  - t' \sum_{{\bf r},\bbox{\gamma},\sigma}
    n^{}_{{\bf r},-\sigma }
    c^\dagger_{{\bf r},\sigma} c^{}_{{\bf r}+\bbox{\gamma},\sigma}
    (1 - n^{}_{{\bf r}+\bbox{\gamma},-\sigma}) \\
  {\cal H}_{01} & = &
  - t \sum_{{\bf r},\bbox{\delta},\sigma}
    (1 - n^{}_{{\bf r},-\sigma})
    c^\dagger_{{\bf r},\sigma} c^{}_{{\bf r}+\bbox{\delta},\sigma}
    n^{}_{{\bf r}+\bbox{\delta},-\sigma} \nonumber \\
  & &
  - t' \sum_{{\bf r},\bbox{\gamma},\sigma}
    (1 - n^{}_{{\bf r},-\sigma})
    c^\dagger_{{\bf r},\sigma} c^{}_{{\bf r}+\bbox{\gamma},\sigma}
    n^{}_{{\bf r}+\bbox{\gamma},-\sigma} \\
  {\cal H}_{00} & = &
  - t \sum_{{\bf r},\bbox{\delta},\sigma}
    (1 - n^{}_{{\bf r},-\sigma})
    c^\dagger_{{\bf r},\sigma} c^{}_{{\bf r}+\bbox{\delta},\sigma}
    (1 - n^{}_{{\bf r}+\bbox{\delta},-\sigma}) \nonumber \\
  & &
  - t' \sum_{{\bf r},\bbox{\gamma},\sigma}
    (1 - n^{}_{{\bf r},-\sigma})
    c^\dagger_{{\bf r},\sigma} c^{}_{{\bf r}+\bbox{\gamma},\sigma}
    (1 - n^{}_{{\bf r}+\bbox{\gamma},-\sigma})
\end{eqnarray}
The notation indicates that each nearest-neighbor pair and
next-nearest-neighbor pair is counted once and once only in the sum.

Writing the result out in its complete form, we have
\begin{eqnarray}
  {\cal H}_{\text{eff}} & = & 
    - t \sum_{{\bf r},\bbox{\delta},\sigma}
      c_{{\bf r},\sigma}^\dagger c_{{\bf r}+\bbox{\delta},\sigma}^{}
    - t' \sum_{{\bf r},\bbox{\gamma},\sigma}
      c_{{\bf r},\sigma}^\dagger c_{{\bf r}+\bbox{\gamma},\sigma}^{}
  \nonumber \\
  & &
    + \frac{J}{2} \sum_{{\bf r},\bbox{\delta}} \left(
      {\bf S}_{\bf r}^{} \cdot {\bf S}_{{\bf r}+\bbox{\delta}}^{}
      -\frac{1}{4} n_{\bf r}^{} n_{{\bf r}+\bbox{\delta}}^{} \right) 
  \nonumber \\
  & &
    - \frac{J}{4} \sum_{{\bf r},\bbox{\delta}'\neq\bbox{\delta},\sigma}
      \left( c_{{\bf r}+\bbox{\delta},\sigma}^\dagger
       n_{{\bf r},-\sigma}^{} c_{{\bf r}+\bbox{\delta}',\sigma}^{} 
      \right.
  \nonumber \\
  & & ~~~~~~~~~~~~~ \left.
     + c_{{\bf r}+\bbox{\delta},\sigma}^\dagger
       c_{{\bf r},-\sigma}^\dagger c_{{\bf r}+\bbox{\delta}',-\sigma}^{}
       c_{{\bf r},\sigma}^{} \right).
  \label{eq:hameff}
\end{eqnarray}
This rather complicated Hamiltonian can be separated into three
physically quite distinct parts.  The first term is  a hoping term of
the usual kind.  An electron hops from the site $ {\bf
r}+\bbox{\delta} $ to the site ${\bf r}$. The only point which must be
borne in mind is that  no doubly-occupied sites can be created in this
process. This kinetic energy term is formally of order $t$.  This is
the leading term in an expansion  in the small parameter $t/U$. 
However, the constraint of no doubly-occupied sites means that only
``holes'' (vacant sites) can move.  Thus, the contribution of this
term to the  energy is proportional to the doping level.  If we call
the  density of vacant sites $\delta$, then the total contribution  of
this term to the energy is of order $z \delta t$,  where $z$ is the
coordination number. ($z=4$ for the square lattice).     The next term
is a spin-spin interaction.  It lowers the energy of antiferromagnetic
spin configurations. The contribution of this term to the total energy
is proportional to $z (1-\delta) J \sim  4 z (1-\delta) t^2 / U$.  The
final group of terms consists of the three-site terms, which represent
a kind of induced kinetic energy through virtual processes. Their
contribution to the energy is of order $z^2 \delta J$.  In the
high-$T_c$ materials, we normally have $J/t \approx 1/3$, and $\delta$
ranges from $0.0$ to $0.4$.  As a function of $\delta$, therefore, we
expect the spin-spin interaction to be the dominant term at very small
$\delta$, while at moderate values of $\delta$, perhaps $\delta
\approx 0.1$, the other two terms  will start to become important,
based on this simple analysis. In fact, which terms are important
depends a good deal on  the property of the system in which we are
interested.


\section{The Dilute Limit}
\label{sec:dilute}

The  $t$-$J$ model has strong support as a description of the
low-lying energy states of the doped CuO$_2$ planes in the
high-temperature superconductors.  Such  an important model must be
investigated over its full range of parameters, not only for the sake
of understanding a model which is interesting in and of itself, but
also for the purpose of gaining insight  into the various trends of
the model as its many variables change from one  area of parameter
space to another.  In other words, such a wide-range  investigation of
the $t$-$J$ model could link nonphysical, yet soluble  limits of the
model to more physical, insoluble regions.  In this spirit of 
identifying trends, we present in this section an analytical solution
to the dilute or low spin-density limit.  The basic $t$-$J$ model will 
first be considered, followed by the model with the addition of
three-site terms \cite{gjr} and  next-nearest-neighbor hopping.
Justification for this ``extended'' $t$-$J$  model is made below.

\subsection{The basic $t$-$J$ model}
\label{sec:dilutea}

Consider two spins, one up and one down, on a square, two-dimensional 
lattice of $N_s$ sites.  The goal is to find the ground state of the 
system satisfying the Schr\"{o}dinger equation 
\begin{equation}
  \tilde{\cal H}_{tJ} \left| \psi \right> = E \left| \psi \right>
  \label{eq:schr}, 
\end{equation} 
where $E$ is the binding energy of the two spins. These spins interact
as described by the ``traditional'' $t$-$J$ model
Eq.~(\ref{eq:hameff}) which may be written as:
\begin{eqnarray}
  {\cal H}_{tJ}^{} & = &
    - t \sum_{{\bf r},\bbox{\delta},\sigma}
    c_{{\bf r},\sigma}^\dagger c_{{\bf r}+\bbox{\delta},\sigma}^{}
    \label{eq:tJham} \\
  & &
    - \frac{J}{4} \sum_{{\bf r},\bbox{\delta},\sigma}
    (n_{{\bf r},-\sigma}^{} n_{{\bf r}+\bbox{\delta},\sigma}^{}
    + c_{{\bf r}+\bbox{\delta},\sigma}^\dagger c_{{\bf r},-\sigma}^\dagger
      c_{{\bf r}+\bbox{\delta},-\sigma}^{} c_{{\bf r},\sigma}^{}), \nonumber
\end{eqnarray}
where summations are over the $N_s$ sites ${\bf r}$, the four
nearest-neighbor sites $\bbox{\delta}$, and spin up and down $\sigma$.
The $J$-term is written in the above form for convenience  in
comparing it with additional terms later. To find the ground state of
the system, the following pair wavefunction form is used:
\begin{equation}
  \left| \psi \right> = \sum_{{\bf r}_1,{\bf r}_2}
     a({\bf r}_1^{}-{\bf r}_2^{})   
     c_{{\bf r}_1\uparrow}^\dagger
     c_{{\bf r}_2\downarrow}^\dagger \left| 0 \right>
  \label{eq:wavefun},
\end{equation}
where the solution entails finding $a({\bf r})$ at all ${\bf r}$. 
Before  this is done, it should be noted that, although the $t$-$J$
model implicitly  includes the restriction of no-doubly-occupied
sites, the given form in Eq.~(\ref{eq:tJham}) will not reflect this
constraint unless an on-site repulsion term is added as
\begin{eqnarray}
  \tilde{\cal H}_{tJ} & = & {\cal H}_{tJ} + {\cal H}_{V_{0}}, \nonumber \\
  {\cal H}_{V_{0}} & = &
  V_{0} \sum_{\bf r} n_{{\bf r}\uparrow} n_{{\bf r}\downarrow} 
  \label{eq:tJVham}, 
\end{eqnarray}
where the limit $V_{0} \rightarrow \infty$ will yield the desired 
constraint $a({\bf r} = 0) = 0$.  In other words, the probability of
finding the pair on the same site will be zero as required.  This
added  term is, of course, identical to the Hubbard $U$-term of
Eq.~(\ref{eq:hubbU}).

Substitution of Eqs.~(\ref{eq:wavefun}) and (\ref{eq:tJVham}) into 
Eq.~(\ref{eq:schr}) and appropriate relabeling of summation variables
reduce the problem to the following equation in $a$:
\begin{eqnarray}
  -2t \sum_{\bbox{\delta}} a({\bf r}+\bbox{\delta})
  -\frac{J}{2} \sum_{\bbox{\delta}}
  (a({\bf r}) - a({\bf r}-2\bbox{\delta}))
  \delta_{{\bf r},\bbox{\delta}}
  \nonumber \\
  + V_{0} a({\bf r}) \delta_{{\bf r},0}
  = E a({\bf r})
\label{eq:aeq} ,
\end{eqnarray}
where ${\bf r}_1$ and ${\bf r}_2$ are fixed, and the relative
coordinate ${\bf r} = {\bf r}_1 - {\bf r}_2$ is used.  Using the
following  relations:
\begin{eqnarray}
a({\bf k}) & = & \frac{1}{N_s}
  \sum_{\bf r} e^{-i{\bf k}\cdot{\bf r}} a({\bf r}), \\
a({\bf r}) & = &
  \sum_{\bf k} e^{i{\bf k}\cdot{\bf r}} a({\bf k}), 
\end{eqnarray}
the Fourier transform of  Eq.~(\ref{eq:aeq}) is taken, and solving for
$a({\bf k})$ yields
\begin{eqnarray}
a({\bf k}) & = & \frac{J}{2 N_s} \sum_{\bbox{\delta}}
     a(\bbox{\delta})
     \frac{\cos{{\bf k} \cdot \bbox{\delta}}}{\varepsilon({\bf k}) - E/2}
     \nonumber \\
     & & - \frac{V_{0}}{2 N_s}a(0)\frac{1}{\varepsilon({\bf k}) - E/2}
\label{eq:ak},
\end{eqnarray}
where $\varepsilon({\bf k}) = -2t(\cos k_{x} + \cos k_{y})$ is the 
tight-binding energy dispersion.  The inverse Fourier transform of 
Eq.~(\ref{eq:ak}) is then taken, resulting in the following:
\begin{eqnarray}
a({\bf r}) & = & \frac{J}{2} \sum_{\bbox{\delta}} a(\bbox{\delta})
  \frac{1}{N_s} \sum_{\bf k}
  \frac{e^{i{\bf k}\cdot{\bf r}} \cos{{\bf k} \cdot \bbox{\delta}}}
  {\varepsilon({\bf k}) - E/2} \nonumber \\
  & & - \frac{V_{0}}{2} a(0)
  \frac{1}{N_s} \sum_{\bf k}
  \frac{e^{i{\bf k}\cdot{\bf r}}}{\varepsilon({\bf k}) - E/2}.
\label{eq:ar1}
\end{eqnarray}
$a(0)$ can be found self-consistently from Eq.~(\ref{eq:ar1}) to be
\begin{eqnarray}
a(0) & = & \frac{J}{V_{0}} \frac{I_{1}}{I_{0} + 2/V_{0}}
  \sum_{\bbox{\delta}} a(\bbox{\delta}), \\
I_{0} & \equiv & \frac{1}{N_s} \sum_{\bf k}
  \frac{1}{\varepsilon({\bf k}) - E/2},
\label{eq:i0def} \\
I_{1} & \equiv & \frac{1}{N_s} \sum_{\bf k}
  \frac{\cos{k_{x}}}{\varepsilon({\bf k}) - E/2}.
\label{eq:i1def}
\end{eqnarray}
Substituting the solution of $a(0)$ back into Eq.~(\ref{eq:ar1}) then
yields the following:
\begin{eqnarray}
a({\bf r}) & = & \frac{J}{2} \sum_{\bbox{\delta}} a(\bbox{\delta})
  \nonumber \\
  & & \times
  \frac{1}{N_s} \sum_{\bf k}
  \frac{e^{i{\bf k}\cdot{\bf r}}
  \left( \cos{{\bf k} \cdot \bbox{\delta}}
  - I_{1}/(I_{0} + 2/V_{0}) \right) }
  {\varepsilon({\bf k}) - E/2}.
\label{eq:ar2}
\end{eqnarray}
It can be seen readily at this point that $a(-{\bf r}) = a({\bf r})$
as  should be expected from the symmetric form of the wavefunction
chosen in  Eq.~(\ref{eq:wavefun}) above.  The four nearest-neighbor
$a(\bbox{\delta})$'s can now be found.  Letting $\bbox{\delta}_{1} =
(1,0)$, $\bbox{\delta}_2 = (0,1)$, $\bbox{\delta}_{3} = (-1,0)$, and
$\bbox{\delta}_{4} = (0,-1)$, the even symmetry of  $a({\bf r})$
requires that $a(\bbox{\delta}_{3}) = a(\bbox{\delta}_{1})$ and
$a(\bbox{\delta}_{4}) = a(\bbox{\delta}_2)$.  Solving for both
$a(\bbox{\delta}_{1})$ and $a(\bbox{\delta}_2)$ self-consistently in
Eq.~(\ref{eq:ar2}) results in  $a(\bbox{\delta}_2) = \pm
a(\bbox{\delta}_{1})$.  Thus, two  solutions for $a({\bf r})$ are
obtained:
\begin{eqnarray}
a_s({\bf r}) & = & J a(\bbox{\delta}_{1})
  \frac{1}{N_s} \sum_{\bf k} e^{i{\bf k}\cdot{\bf r}} \nonumber \\
  & & \times \frac{
  \left( \cos{k_{x}} + \cos{k_{y}} - 2 I_{1}/(I_{0} + 2/V_{0}) 
  \right)}
  {\varepsilon({\bf k}) - E/2}
\label{eq:sar} \\
a_{d}({\bf r}) & = & J a(\bbox{\delta}_{1})
  \frac{1}{N_s} \sum_{\bf k}
  e^{i{\bf k}\cdot{\bf r}}
  \frac{
  \left( \cos{k_{x}} - \cos{k_{y}} \right) }
  {\varepsilon({\bf k}) - E/2},
\label{eq:dar}
\end{eqnarray}
where Eqs.~(\ref{eq:sar}) and (\ref{eq:dar}) correspond to an extended 
$s$-wave ($s_{x^2+y^2}$) solution and a $d$-wave ($d_{x^2-y^2}$)
solution, respectively.

The $d_{x^2-y^2}$-wave solution in Eq.~(\ref{eq:dar}) has no
dependence on  $V_{0}$, and it can be seen by symmetry alone that
$a_{d}(0)=0$; that is, there is  no double-occupancy.  The  $s$-wave
solution in Eq.~(\ref{eq:sar})  also obtains this constraint upon
letting the on-site repulsion $V_{0}$  become infinite:
\begin{eqnarray}
\lim_{V_{0} \rightarrow \infty} a_s({\bf r}) & = &
  J a(\bbox{\delta}_{1}) \frac{1}{N_s} \sum_{\bf k}
  e^{i{\bf k}\cdot{\bf r}} \nonumber \\
  & & \times \frac{
  \left( \cos{k_{x}} + \cos{k_{y}} - 2 I_{1}/I_{0} \right)}
  {\varepsilon({\bf k}) - E/2},
\end{eqnarray}
where it can be seen that $a_s(0) = 0$ by using the definitions of 
$I_{0}$ and $I_{1}$ in Eqs.~(\ref{eq:i0def}) and (\ref{eq:i1def})
above. Finally, in the thermodynamic limit ($N_s \rightarrow \infty$),
the sums over ${\bf k}$ become integrals over $d{\bf k}$, and the
solutions to the ``traditional'' $t$-$J$ model in the dilute limit of
one pair of spins become
\begin{eqnarray}
a_s({\bf r}) & = & J a(\bbox{\delta}_{1})
  \frac{1}{(2 \pi)^2} \int d{\bf k}
  e^{i{\bf k}\cdot{\bf r}} \nonumber \\
  & & \times \frac{
  \left( \cos{k_{x}} + \cos{k_{y}} - 2 I_{1}/I_{0} \right)}
  {\varepsilon({\bf k}) - E/2},
\label{eq:asr} \\
a_{d}({\bf r}) & = & J a(\bbox{\delta}_{1})
  \frac{1}{(2 \pi)^2} \int d{\bf k}
  e^{i{\bf k}\cdot{\bf r}} \nonumber \\
  & & \times \frac{
  \left( \cos{k_{x}} - \cos{k_{y}} \right)}
  {\varepsilon({\bf k}) - E/2},
\label{eq:adr} \\
I_{0} & = & \frac{1}{(2 \pi)^2} \int d{\bf k}
  \frac{1}{\varepsilon({\bf k}) - E/2} ,
\label{eq:I0} \\
I_{1} & = & \frac{1}{(2 \pi)^2} \int d{\bf k}
  \frac{\cos{k_{x}}}{\varepsilon({\bf k}) - E/2},
\label{eq:I1}
\end{eqnarray}
where integration is over the full Brillouin zone: $k_{x} = -\pi$ to
$\pi$  and $k_{y} = -\pi$ to $\pi$.

Of some interest is the critical value of $J$ at which there is a
bound state for  each solution.  Letting ${\bf r} = \bbox{\delta}_{1}$
in Eqs.~(\ref{eq:asr}) and (\ref{eq:adr}), the relationships between
$J$ and $E$ for the two symmetries can be obtained:
\begin{eqnarray}
\frac{1}{J_s}  =  \frac{1}{(2 \pi)^2} \int d{\bf k}
  \frac{\cos{k_{x}}
  \left( \cos{k_{x}} + \cos{k_{y}} - 2 I_{1}/I_{0} \right)}
  {\varepsilon({\bf k}) - E/2},  
\label{eq:jcs} \\
\frac{1}{J_{d}}  =  \frac{1}{(2 \pi)^2} \int d{\bf k}
  \frac{\cos{k_{x}}
  \left( \cos{k_{x}} - \cos{k_{y}} \right)}
  {\varepsilon({\bf k}) - E/2}
\label{eq:jcd}.
\end{eqnarray}
These integrals cannot be done in closed form but lend themselves
fairly  easily to straightforward numerical methods such as Monte
Carlo  integration.  Some care must be taken when considering small
$E$ where the  numerator tends to zero at ${\bf k} = 0$.  In the $J
\rightarrow \infty$ limit, there are precisely four  bound states
corresponding to the four sites on which the  attraction is nonzero. 
Two of these are $p$-wave, and therefore not allowable.  Thus there
are only two bound state solutions in this limit.  Since the number of
bound states can only decrease as the attraction is decreased, the
maximum number of bound states is two.  The  $s$-wave and $d$-wave
solutions discussed here are the only possible bound states in the
model. 

Equations (\ref{eq:jcs}) and (\ref{eq:jcd}) are the basic equations
determining the bound state energies, and we derived them from the
Schr\"odinger equation. However, we may derive them from the BCS gap
equation as well.  In this regard, see also Refs. \cite{leg,jap,rand}.
At zero temperature the gap equation is
\begin{equation}
\Delta({\bf k}) = - \sum_{\bf k'} V({\bf k},{\bf k'})
\frac{\Delta({\bf k'})}
{\sqrt{(\varepsilon({\bf k})-\mu)^2 + \Delta({\bf k})^2}}.
\label{eq:gapequ}
\end{equation}
In the limit when $|\Delta({\bf k})|$ is small, this becomes linear:
\begin{equation}
\Delta({\bf k}) = - \sum_{\bf k'} V({\bf k},{\bf k'})
\frac{\Delta({\bf k'})} {|\varepsilon({\bf k})-\mu|}.
\label{eq:stab}
\end{equation}
This is the {\it stability} form of the gap equation.  If it has a
solution, then the system is superconducting (with infinitesimal gap)
at zero temperature.  The reason this form of the gap equation is
rarely seen is that, in conventional superconductivity with BCS-like
purely attractive interactions, superconductivity at zero temperature
occurs even with infinitesimal interaction strength.  Here this is not
necessarily the case. Note also that no frequency cutoff is assumed in
the interaction, because the basic model contains no retardation. 
This is another important  difference from the  usual BCS model.

We may now transform Eq.~(\ref{eq:stab}) into Eq.~(\ref{eq:jcs}) by
first making the substitutions $\mu = 0$, (corresponding to the dilute
case), $\Delta({\bf k}) = \Delta_s (\cos(k_x) + \cos(k_y))$, and
$V({\bf k},{\bf k'}) = -J (\cos(k_x-k_x') + \cos(k_y-k_y'))$, then
multiplying the resulting equation by $\cos(k_x) + \cos(k_y)$ and
integrating over ${\bf k}$.  The result is Eq.~(\ref{eq:jcs}). A
similar procedure using  $\Delta({\bf k}) = \Delta_d (\cos(k_x) -
\cos(k_y))$ yields Eq.~(\ref{eq:jcd}).
Hence the gap equation at zero
gap is the same as the Schr\"odinger equation at zero binding energy.


\begin{figure}
\begin{center}
\leavevmode
\epsffile{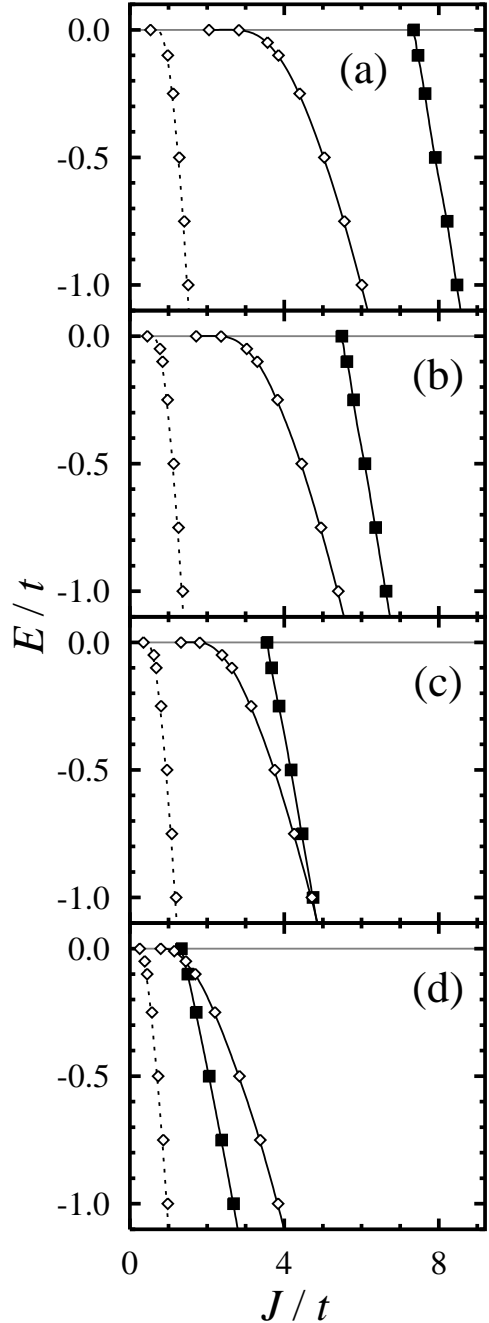}
\caption[]{
Phase diagram of binding energy $E$ vs.\ coupling constant $J$ for the
extended $t$-$J$ model in the dilute limit for (a) $t' = 0.00~t$, (b)
$t' = -0.15~t$, (c) $t' = -0.30~t$ and (d) $t' = -0.45~t$.  The solid
line plots correspond to calculations done without the three-site term
($J_3 = 0$), the dashed lines for those done with the three-site term
($J_3 = J$), the plots with open diamonds are for $s_{x^2+y^2}$-wave,
and the plots with closed squares are for $d_{x^2-y^2}$-wave.  The
addition of  the $J_3$-term causes the $s$-wave curves to shift left,
whereas the $d$-wave curves shift right off to infinity;  that is,
there is no  $d$-wave solution.  The addition of the $t'$-term causes
$d$-wave (for $J_3 = 0$) to shift left and cross the $s$-wave curve at
a point dependent upon the strength of $t'$.}
\label{fig:phdiag}
\end{center}
\end{figure}

A phase diagram of $E$ vs.~$J$ depicting the  $s$- and $d$-wave bound
states may be seen in Fig.~\ref{fig:phdiag}(a).  It can be noted from
this low density phase diagram that  $s$-wave always has the  lower
energy of the two solutions at fixed $J$ and is, therefore, the ground
state of  the problem at hand.  Furthermore,  $s$-wave has a bound
state  above a critical-$J$ ($J_{c}$) of $2.00~t$ whereas $d$-wave has
a $J_{c}$  of about $7.32~t$. The phase diagram of
Fig.~\ref{fig:phdiag}(a) is  identical to that reported recently by
Hellberg and  Manousakis \cite{hellberg} and the $s$-wave results are
also similar to those of Kagan and Rice \cite{kagan}.

If one holds to the idea that magnetic pairing should produce
$d$-wave, because of the hard core in the potential, the  surprising
aspect of these results is that $s$-wave always has by far the largest
binding energy of the two symmetries.  We may in fact prove a general
theorem that if  the potential has square symmetry, the ground state
is always $s$-wave for nearest-neighbor hopping.  This theorem holds
whether there is a hard core or not.  The proof is given first on the
continuum with a radially symmetric potential,  since it illustrates
the main ideas most clearly. The Hamiltonian is
\begin{equation}
{\cal H} = -\frac{\hbar^2}{2 \mu} 
\left[\frac{1}{\rho}\frac{\partial}{\partial \rho} 
\left(\rho \frac{\partial}{\partial \rho}\right) + 
\frac{1}{\rho^2} \frac{\partial^2}{\partial \phi^2}\right] + V(\rho).
\end{equation}
We argue by contradiction.  Let $\psi_m(\rho,\phi)$ be the normalized
ground state and have non-$s$-wave symmetry:
\begin{equation}
\psi_m(\rho,\phi) = \frac{1}{\sqrt{2 \pi}}f(\rho)e^{im\phi},
\end{equation}        
with $m \neq 0$ and $\int_0^\infty \rho f^2(\rho) d \rho = 1$.  The
ground state energy is
\begin{eqnarray}
E_m & = & \left<\psi_m|{\cal H}|\psi_m\right> \nonumber \\
& = &
-\frac{\hbar^2}{2 \mu} 
\int_0^\infty f(\rho) \frac{\partial}{\partial \rho} 
\left(\rho \frac{\partial f(\rho)}{\partial \rho}\right) d \rho \nonumber \\
& & + \frac{m^2 \hbar^2}{2 \mu} \int_0^\infty
\frac{1}{\rho} f^2(\rho) d \rho \nonumber \\
& & + \int_0^\infty \rho f^2(\rho) V(\rho) d \rho.
\end{eqnarray}
Now consider the normalized trial wavefunction  $\psi_s(\rho,\phi) =
f(\rho)$, which is $s$-wave.  This has the expectation value
\begin{eqnarray}
E_s & = & \left<\psi_s|{\cal H}|\psi_s\right> \nonumber \\
& = &
-\frac{\hbar^2}{2 \mu} 
\int_0^\infty f(\rho) \frac{\partial}{\partial \rho} 
\left(\rho \frac{\partial f(\rho)}{\partial \rho}\right)
d \rho \nonumber \\
& & + \int_0^\infty \rho f^2(\rho) V(\rho) d \rho.
\end{eqnarray}
Clearly,
\begin{equation}
E_s < E_d = E_s + 
\frac{m^2 \hbar^2}{2 \mu} \int_0^\infty
\frac{1}{\rho} f^2(\rho) d \rho. 
\end{equation}
Thus $\psi_m$ cannot be the ground state.  A similar proof obviously 
holds in three dimensions.  The physics is simply that  the increased
kinetic energy from angular variations in a higher-wave wavefunction
will always be greater than any gain in potential energy coming from
avoidance of the core.

On the lattice, the physics is clearly the same, and the proof is
quite similar.  Let the potential $V({\bf r})$ have square symmetry,
and let $ t_{{\bf r}, {\bf r'}} $, the hopping  coefficients, be
positive.  Let the assumed ground state be $\psi_d({\bf r})$ have 
$d$-wave ($B_1$) symmetry.  It may be taken to be real.   Now consider
the wavefunction
\begin{equation}
\psi_s({\bf r}) = |\psi_d({\bf r})|. 
\end{equation}
The expectation values are:
\begin{eqnarray}
E_d & = & \left<\psi_d|{\cal H}|\psi_d\right> \nonumber \\
& = &
\sum_{\bf r} V({\bf r}) \psi_d({\bf r})^2 -
\sum_{{\bf r} \neq {\bf r'}} t_{{\bf r}, {\bf r'}}
\psi_d({\bf r}) \psi_d({\bf r'}),  
\label{eq:dvar}
\end{eqnarray}
and 
\begin{eqnarray}
E_s & = & \left<\psi_s|{\cal H}|\psi_s\right> \nonumber \\
& = & 
\sum_{\bf r} V({\bf r}) \psi_d({\bf r})^2 -
\sum_{{\bf r} \neq {\bf r'}} 
t_{{\bf r}, {\bf r'}} |\psi_d({\bf r}) \psi_d({\bf r'})|.  
\label{eq:svar}
\end{eqnarray}
Comparison shows that
\begin{equation}
E_s \leq E_d,
\label{eq:ineq}
\end{equation}
so that $\psi_d$ can never be the nondegenerate  ground state.  Except
for pathological cases, it is not hard to improve $\psi_s$ to get a
strict inequality  in Eq.~\ref{eq:ineq}. (Let $\psi_s(x,x) \rightarrow
\psi_s(x,x) + \delta \psi$, with $\delta$ sufficiently small).  We
have therefore shown that $s$-wave is in fact always the lowest energy
solution.  Hard core or not, $d$-wave never wins.

\subsection{The addition of the three-site terms}

We now include the so-called three-site terms of the $t$-$J$ model in
our  dilute limit investigations.  These added terms originate from
the  derivation of the $t$-$J$ model from the Hubbard model. 
Considering the  Hubbard model,
\begin{eqnarray}
{\cal H}_{\text{Hubbard}}^{} = -t \sum_{{\bf r},\bbox{\delta},\sigma}
     c_{{\bf r},\sigma}^\dagger c_{{\bf r}+\bbox{\delta},\sigma}^{}
     + U \sum_{\bf r} n_{{\bf r}\uparrow}^{}
                      n_{{\bf r}\downarrow}^{},
\end{eqnarray}
the $t$-term is treated as a perturbation with respect to the
$U$-term,  where $U \gg t$.  Only terms up to and including order
$t^{2}/U$ are  retained.  Letting $J \equiv 4t^{2}/U$, the
perturbative expansion yields the  $t$-$J$ Hamiltonian of
Eq.~(\ref{eq:tJham}) plus the three site terms \cite{gjr}. The
Hamiltonian may now be written
\begin{eqnarray}
  {\cal H}_{tJJ_3}^{} & = & 
    -t \sum_{{\bf r},\bbox{\delta},\sigma} c_{{\bf r},\sigma}^\dagger
    c_{{\bf r}+\bbox{\delta},\sigma}^{} \label{eq:tJJ3ham} \\
  & &
    -\frac{J}{4} \sum_{{\bf r},\bbox{\delta},\sigma}
    (n_{{\bf r},-\sigma}^{} n_{{\bf r}+\bbox{\delta},\sigma}^{}
    + c_{{\bf r}+\bbox{\delta},\sigma}^\dagger c_{{\bf r},-\sigma}^\dagger
    c_{{\bf r}+\bbox{\delta},-\sigma}^{} c_{{\bf r},\sigma}^{}) \nonumber \\
  & & -\frac{J_3}{4}
    \sum_{{\bf r},\bbox{\delta'}\neq\bbox{\delta},\sigma}
    (c_{{\bf r}+\bbox{\delta},\sigma}^\dagger n_{{\bf r},-\sigma}^{}
     c_{{\bf r}+\bbox{\delta'},\sigma}^{} \nonumber \\
  & & ~~~~~~~~~~~~~~
    + c_{{\bf r}+\bbox{\delta},\sigma}^\dagger c_{{\bf r},-\sigma}^\dagger
    c_{{\bf r}+\bbox{\delta'},-\sigma}^{} c_{{\bf r},\sigma}^{}),
\end{eqnarray}
where $J_3$ is merely used as an adjustable factor;  letting $J_3 = 0$ 
restores Eq.~(\ref{eq:tJham}), whereas letting $J_3 = J$ yields the 
extended model as derived from the Hubbard model.

The three-site terms are usually dropped due to their complexity and
the fact that their expectation value is formally of order $\delta J$,
not $J$, where $\delta$ is the hole density. However, they are
enhanced by the fact they are of order $z^2$, not $z$,  where $z$ is
the coordination number.  Numerical results show clearly that  they
become important when $\delta$ is greater than $0.05$
\cite{koltenbah}.   Certainly, it would then be expected in the dilute
limit that the three-site  terms become very important interactions
concerning the ground state energy  of the system, and the solution to
the extended model in the two-spin  limit bears this out.

The solution of two spins interacting on a two-di\-men\-sion\-al
square lattice of  $N_s$ sites is identical to that of
Sec.~\ref{sec:dilutea} and need not be  repeated here as now applied
to Eq.~(\ref{eq:tJJ3ham}).  The wavefunction  coefficients $a({\bf
r})$ of Eq.~(\ref{eq:wavefun}) are found to have the following
solutions (upon letting $N_s \rightarrow \infty$)
\begin{eqnarray}
a_s({\bf r}) & = & J \left( 1 + \frac{3J_3}{J} \right)
  a(\bbox{\delta}_{1})
  \frac{1}{(2 \pi)^2} \int d{\bf k} e^{i{\bf k}\cdot{\bf r}}
  \nonumber \\
  & & \times \frac{
  \left( \cos{k_{x}} + \cos{k_{y}} - 2 I_{1}/I_{0} \right)}
  {\varepsilon({\bf k}) - E/2},
\label{eq:asr2} \\
a_{d}({\bf r}) & = & J \left( 1 - \frac{J_3}{J} \right)
  a(\bbox{\delta}_{1})
  \frac{1}{(2 \pi)^2} \int d{\bf k} e^{i{\bf k}\cdot{\bf r}}
  \nonumber \\
  & & \times \frac{
  \left( \cos{k_{x}} - \cos{k_{y}} \right)}
  {\varepsilon({\bf k}) - E/2}
\label{eq:adr2},
\end{eqnarray}
where $I_{0}$ and $I_{1}$ are still defined as in Eqs.~(\ref{eq:I0})
and (\ref{eq:I1}),  respectively.  As expected, when $J_3 = 0$, the
solutions of the $t$-$J$ model without three-site terms,
Eqs.~(\ref{eq:asr}) and (\ref{eq:adr}), are retrieved.

Given that $J_3 = J$, it can be seen that there is only a trivial
$d$-wave  solution, namely $a_{d}({\bf r}) = 0$ for all ${\bf r}$. In
other words, there  is no bound state solution for $d$-wave pairing. 
The  $s$-wave  solution is similar to that of Eq.~(\ref{eq:asr}),
differing only in  pre-factors.  Finding the relationship between $J$
and $E$ for these two solutions yields
\begin{eqnarray}
\frac{1}{J_s} & = & \left( 1 + \frac{3J_3}{J} \right)
  \frac{1}{(2 \pi)^2} \int d{\bf k} \cos{k_{x}} \nonumber \\
  & & \times \frac{
  \left( \cos{k_{x}} + \cos{k_{y}} - 2 I_{1}/I_{0} \right)}
  {\varepsilon({\bf k}) - E/2},  
\label{eq:jcs2} \\
\frac{1}{J_{d}} & = & \left( 1 - \frac{J_3}{J} \right) 
  \frac{1}{(2 \pi)^2} \int d{\bf k} \cos{k_{x}} \nonumber \\
  & & \times \frac{
  \left( \cos{k_{x}} - \cos{k_{y}} \right)}
  {\varepsilon({\bf k}) - E/2}
\label{eq:jcd2}.
\end{eqnarray}
It can be readily seen that, as the three-site interactions are
``turned on'', that is as $J_3$ is increased from $0$ to $J$, the $J$
vs.~$E$ curve for  $d$-wave would move off to infinity in the phase
diagram of Fig.~\ref{fig:phdiag}(a).  This again shows that there is
no bound state  solution for $d$-wave in the dilute limit of the
extended $t$-$J$ model. For $s$-wave, the $J$  vs.~$E$ curve is
exactly one fourth that of the  $s$-wave solution  for the
``traditional'' $t$-$J$ model as shown in Fig.~\ref{fig:phdiag}(a).
Thus, the three-site terms are $s$-wave enhancing, $d$-wave
suppressing interactions.

The expectation values of the various terms of the extended $t$-$J$
model can  be expressed in terms of the wavefunction coefficients
$a({\bf r})$ as well  as in integral forms.  The normalization is
\begin{eqnarray}
\left< \psi | \psi \right> & = & \sum_{{\bf r}_1,{\bf r}_2}
    |a({\bf r}_1-{\bf r}_2)|^{2} \nonumber \\
    & = &
    J^{2} B_{J_3}^{2} |a(\bbox{\delta}_{1})|^2
    \frac{N_s}{(2 \pi)^{2}}
    \int d{\bf k} A_{k}^{2},
\end{eqnarray}
where $N_s$ is temporarily left finite but will cancel out in the 
expectation values shown below. The coefficients $B_{J_3}$ and $A_{\bf
k}$ are defined for convenience as
\begin{eqnarray}
  A_{\bf k} & = & (\cos{k_{x}} + \cos{k_{y}} - 2 I_{1}/I_{0})/
     (\varepsilon({\bf k}) - E/2), \nonumber \\
  B_{J_3} & = & 1 + 3 J_3/J
\end{eqnarray}
for $s$-wave and
\begin{eqnarray}
  A_{\bf k} & = & (\cos{k_{x}} - \cos{k_{y}})/
     (\varepsilon({\bf k}) - E/2), \nonumber \\
  B_{J_3} & = & 1 - J_3/J
\end{eqnarray}
for $d$-wave.

The expectation values can now be expressed in terms of integrals. 
The  relations between $J$ and $E$ from Eqs.~(\ref{eq:jcs2}) and
(\ref{eq:jcd2})  are also employed to yield
\begin{eqnarray}
\left<{\cal H}_{t}\right> & = & \frac{\left<\psi|{\cal H}_{t}|\psi\right>}
     {\left<\psi|\psi\right>} \nonumber \\
     & = &
     -4t \frac{\int d{\bf k} A_{\bf k}^{2}
     (\cos{k_{x}} + \cos{k_{y}})}
     {\int d{\bf k} A_{\bf k}^{2}}, \\
\left<{\cal H}_{J}\right> & = & \frac{\left<\psi|{\cal H}_{J}|\psi\right>}
     {\left<\psi|\psi\right>} \nonumber \\
     & = &
     -\frac{4}{B_{J_3}} \frac{\int d{\bf k} A_{\bf k} \cos{k_{x}}}
     {\int d{\bf k} A_{\bf k}^{2}}, \\
\left<{\cal H}_{J_3}\right> & = & \frac{\left<\psi|{\cal H}_{J_3}|\psi\right>}
     {\left<\psi|\psi\right>} \nonumber \\
     & = &
     -\frac{12}{B_{J_3}} \left( \frac{J_3}{J} \right)
     \frac {\int d{\bf k} A_{\bf k} \cos{k_{x}}}
     {\int d{\bf k} A_{\bf k}^{2}}, \hspace*{0.1in} \text{$s$-wave} \\
     & = &
     +\frac{4}{B_{J_3}} \left( \frac{J_3}{J} \right)
     \frac {\int d{\bf k} A_{\bf k} \cos{k_{x}}}
     {\int d{\bf k} A_{\bf k}^{2}}, \hspace*{0.1in} \text{$d$-wave}.
\end{eqnarray}
Of course, the solutions for $d$-wave are ill-defined for $J_3 = J$ 
($a_d({\bf k}) = 0$ for all ${\bf k}$ in this case) but are included
here merely for completeness. What can be immediately seen is that,
for  $s$-wave, the three-site  term has an expectation value exactly
three times that of the $J$-term, or $\left<{\cal H}_{J_3}\right> =
3\left<{\cal H}_{J}\right>$, and for $d$-wave, the two expectation
values are equal and opposite, or $\left<{\cal H}_{J_3}\right> =
-\left<{\cal H}_{J}\right>$.


\begin{figure}[ht]
\begin{center}
\leavevmode
\epsffile{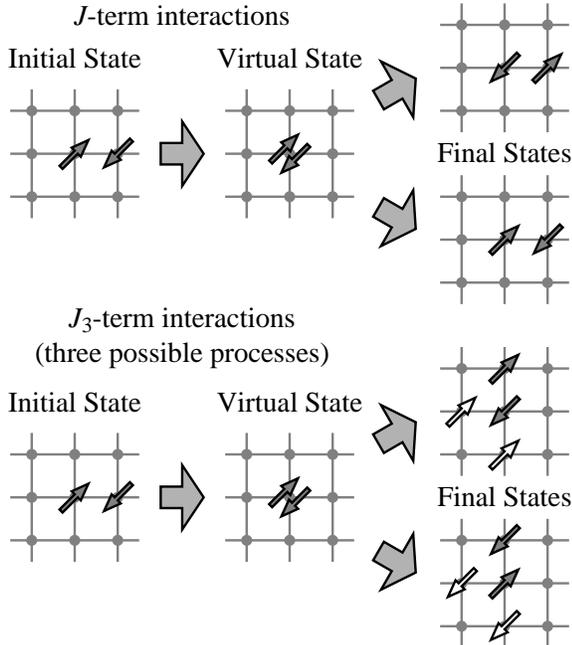}
\caption[]{
Diagram of the dynamics of the spin-spin correlation term (or
$J$-term) and the three-site term (or $J_3$-term).  Shown are initial,
intermediate (or virtual), and final states for one example of each
interaction (filled arrows).  For the $J_3$-term, alternative 
interactions are also shown (open arrows).
In these examples, a down spin has a virtual hopping to a
nearest-neighbor site occupied by an up spin.  In the $J$-term,
either spin hops back to the original site, and in the $J_3$-term,
either spin hops to one of the other three nearest-neighbor sites. 
In the dilute limit, the spins have equal probability of hopping
to any of the four nearest-neighbor sites as shown here.  In the
Cooper problem, however, there would be a probability of $\delta$ for
each of the three originally unoccupied nearest-neighbor sites to
actually be occupied by a non-interacting spin.}
\label{fig:JJ3term}
\end{center}
\end{figure}

A closer look at the various interactions of the extended $t$-$J$
model in  the dilute limit is worth making at this point.  It is
obvious from  the Hamiltonian as written in Eq.~(\ref{eq:tJJ3ham})
that the $J_3$-term is  quite similar to the $J$-term.  The three-site
terms comprise interactions  involving a site and two distinct
nearest-neighbors, whereas the $J$-term is  the special case where the
two nearest-neighbor sites are merely the same  site. Figure
\ref{fig:JJ3term} shows a depiction of the virtual processes involved 
in the spin-spin correlation term ($J$-term) and the three-site terms
($J_3$-term). When the two spins are on nearest-neighbor sites, the
$J$-term can be thought of as a virtual hopping of one spin onto the
site  occupied by the other spin, followed by either spin hopping back
to the  first site.  In the dilute limit, the probability of finding
the other three  nearest-neighbor sites unoccupied is exactly $1$. 
Hence, when one spin has a  virtual hopping onto the site occupied by
the other spin, there is equal  chance of one of the spins hopping
onto any of the four nearest-neighbors.   For  $s$-wave, it is easy to
see then why the $J_3$-term is three  times the size of the $J$-term,
and, due to its change of sign with $\pi/2$  rotation, the $d$-wave
$J_3$-term is equal to the $J$-term but with opposite sign.  Hence,
the total summation of the $J_3$- and $J$-terms for $d$-wave  yields
zero due to its symmetry, and the total summation for   $s$-wave
yields four times the value of the $J$-term. The importance of this
result will become more apparent  in Sec.~\ref{sec:cooper} where
increasing spin density is considered in the Cooper problem.  The
actual effect of the three-site terms is more akin to a kinetic energy
term than an interaction term. It is an effective hopping to second-
and third-nearest-neighbors. It is therefore not so surprising that it
tends to stabilize the  $s$-wave state.

\subsection{The addition of the next-nearest-neighbor hopping term}

One more interaction is included in our extended $t$-$J$ model, namely
the  next-nearest-neighbor hopping term.  This term is added with the
intent of  making a first approximation at modeling specific
high-temperature superconducting systems.  It is known, for example
from neutron scattering, that the various high-$T_{c}$ copper-oxides
differ in energy dispersion, Fermi surface, etc.   Using the
next-nearest-neighbor hopping term with coefficient $t'$ is a way  of
distinguishing these compounds from one another in our tight-binding 
model.  Perhaps a more systematic approach, emulating our derivation
of the  $J_3$-term, would be to begin with the Hubbard model and
include the  $t'$-term in the large-$U$ perturbative expansion.  This
would result in  more terms of order $J$, that is various three-site
terms involving  nearest-neighbor sites, next-nearest-neighbor sites,
and mixtures of  both.  The many extra terms may be worth studying at
some future time, but for  the purposes of identifying trends in the
$t$-$J$ model, we do not wish to  introduce even more complexity to
our extended model until it has been  sufficiently investigated in its
present form.

The extended $t$-$J$ model is given in its full form as
\begin{eqnarray}
  {\cal H}_{tt'JJ_3}^{} & = & 
    -t \sum_{{\bf r},\bbox{\delta},\sigma}
    c_{{\bf r},\sigma}^\dagger c_{{\bf r}+\bbox{\delta},\sigma}^{}
    -t' \sum_{{\bf r},\bbox{\gamma},\sigma}
    c_{{\bf r},\sigma}^\dagger c_{{\bf r}+\bbox{\gamma},\sigma}^{}
    \label{eq:tt2JJ3ham} \\
  & &
    -\frac{J}{4} \sum_{{\bf r},\bbox{\delta},\sigma}
    (n_{{\bf r},-\sigma}^{} n_{{\bf r}+\bbox{\delta},\sigma}^{}
    \nonumber \\
  & & ~~~~~~~~~
    + c_{{\bf r}+\bbox{\delta},\sigma}^\dagger
    c_{{\bf r},-\sigma}^\dagger
    c_{{\bf r}+\bbox{\delta},-\sigma}^{}
    c_{{\bf r},\sigma}^{}) \nonumber \\
  & &
    -\frac{J_3}{4}
    \sum_{{\bf r},\bbox{\delta'}\neq\bbox{\delta},\sigma}
    (c_{{\bf r}+\bbox{\delta},\sigma}^\dagger
    n_{{\bf r},-\sigma}^{} c_{{\bf r}+\bbox{\delta'},\sigma}^{}
    \nonumber \\
  & & ~~~~~~~~~~~~~~
    + c_{{\bf r}+\bbox{\delta},\sigma}^\dagger
    c_{{\bf r},-\sigma}^\dagger
    c_{{\bf r}+\bbox{\delta'},-\sigma}^{}
    c_{{\bf r},\sigma}^{}), \nonumber
\end{eqnarray}
where summation over $\bbox{\gamma}$ means summation over 
next-nearest-neighbor sites.  Once again considering the problem in
the  dilute limit, the solution to the problem of two spins
interacting with the  Hamiltonian of Eq.~(\ref{eq:tt2JJ3ham}) results
in exactly the same forms for  $a_s({\bf r})$ and $a_{d}({\bf r})$ as
given in Eqs.~(\ref{eq:asr2}) and (\ref{eq:adr2}), respectively, only
now, the tight-binding energy dispersion is given by $\varepsilon({\bf
k}) = -2t(\cos{k_{x}} + \cos{k_{y}}) -  4t'\cos{k_{x}}\cos{k_{y}}$.

The relationships between $J$ and $E$ are of the  same forms as given
in Eqs.~(\ref{eq:jcs2}) and (\ref{eq:jcd2}) for  $s$-wave and
$d$-wave, respectively.  Once again, phase diagrams for $E$  vs.~$J$
can be drawn for various values of $t'$.  Fig.~\ref{fig:phdiag}(a) 
is, obviously, the phase diagram with $t' = 0$. 
Figs.~\ref{fig:phdiag}(b),  (c), and (d) show the phase diagrams for
$t' = -0.15~t$, $-0.30~t$, and  $-0.45~t$, respectively.  It can be
seen that, for finite $t'$ and no  three-site interactions, the  
$s$-wave and $d$-wave curves cross at a binding energy $E$ and value
$J$ which decrease with increasing magnitude of $t'$.  In other words,
as the  next-nearest-neighbor hopping term increases in strength,
there is a  cross-over between an $s$-wave ground state solution and a
$d$-wave ground  state solution at smaller and smaller binding
energies and $J$-values.  This shows that the $t'$-term with a
negative sign is an $s$-wave suppressing, $d$-wave enhancing term in
the model.   That this must be the case is evident already from
Eqs.~(\ref{eq:dvar}) and (\ref{eq:svar}).  Once again, when the
three-site terms are considered  ($J_3 = J$), there is no $d$-wave
solution in the dilute limit, and the  $s$-wave curve is reduced to a
quarter of its value without the  $J_3$-term.

The physics of the dilute limit is that, in the absence of a Fermi
surface, the $s$-wave instability is dominant.  The $s$-wave  simply
has lower kinetic energy.  In the next section we present an 
investigation of a single pair now interacting in the presence of a 
Fermi surface: the extended $t$-$J$ model is evaluated in the  Cooper
problem.


\section{The Cooper Problem}
\label{sec:cooper}

The investigations of Sec.~\ref{sec:dilute} can now be extended  from
the dilute limit to the regime of finite density.  Again, the  purpose
of this study is to identify trends with the addition of the
three-site term, with the changing of the next-nearest-neighbor
hopping term, and now with the changing of the density of spins.  As
before, the  problem is solved for one pair of interacting spins, now
being in the presence of a filled Fermi sea of non-interacting spins.
These results will be compared qualitatively  with previous VMC
results which also yield some of the trends detailed in this work.

Consider the problem again of two spins interacting on a square,
two-dimensional lattice of $N_s$ sites, but now allow them to interact
in the presence of a filled Fermi sea up to a Fermi energy (chemical
potential) of $\mu$.  This is the Cooper problem as applied to the
extended $t$-$J$ model.  As we shall show, this also corresponds again
to the stability form of the gap equation at zero temperature with a
certain interaction.

The solution once again entails finding the wavefunction coefficients
$a({\bf r})$ Eq.~(\ref{eq:wavefun}) from the Schr\"{o}dinger equation
Eq.~(\ref{eq:schr}) which is now rewritten
\begin{equation}
\tilde{\cal H}_{tt'JJ_3} \left| \psi \right> = (2\mu + E)
  \left| \psi \right>
\label{eq:schr2},
\end{equation}
where the eigenvalue $E$ has been redefined with respect to the Fermi
energy. Once again, $\tilde{\cal H}_{tt'JJ_3}$ must include an on-site
repulsion term so that the solution will enforce no-double-occupancy
as was shown in Eq.~(\ref{eq:tJVham}).  In a similar fashion, the
states within the Fermi surface must be excluded from any interactions
since they are to be filled with non-interacting spins.  This is still
a two-particle problem which reduces to a single-particle equation as
in Eq.~(\ref{eq:aeq}), but with a fictitious interaction coming from 
the Pauli exclusion principle.  The exclusion of states within the
Fermi surface is taken into account by adding another term to
$\tilde{\cal H}_{tt'JJ_3}$, namely
\begin{equation}
\tilde{\cal H}_{tt'JJ_3} \rightarrow \tilde{\cal H}_{tt'JJ_3}
+ \sum_{{\bf k},\sigma} U_C({\bf k}) n_{{\bf k},\sigma}
\end{equation}
with the following dependence on ${\bf k}$:
\begin{equation}
U_C({\bf k}) = \left\{
\begin{array}{ll}
U_C^0, & \xi({\bf k}) < 0 \\
0, & \xi({\bf k}) > 0
\end{array}
\right.,
\label{eq:Vk}
\end{equation}
where $\xi({\bf k}) = \varepsilon({\bf k}) - \mu$.  Thus, the same
technique as with the exclusion of doubly-occupied sites can be
employed:  the limit of $U_C^0 \rightarrow \infty$ at the end of the
derivation will exclude states within the Fermi surface.
       
An important observation  needs to be made concerning the three-site
terms.  Figure \ref{fig:JJ3term}  shows the dynamics of the $J$-term
and $J_3$-term in the dilute limit,  where the $J_3$-term  has three
times the value of the $J$-term for extended $s$-wave and  the same
value and opposite sign for $d$-wave.  This is due to the fact that,
if the two spins occupy nearest-neighbor sites, the other three
nearest-neighbor sites would then be guaranteed unoccupied in the
dilute limit. Now, however, with finite density,  if the two
interacting spins are on nearest-neighbor sites, then there would be a
virtual hopping onto one of the sites, and the probability of the
original  nearest-neighbor being unoccupied would be $1$, whereas the
probability of  the other three nearest-neighbors being unoccupied
would be $\delta$, the  hole density.  In other words, the three-site
terms need to have a  pre-factor of $\delta$ to take into account the
finite density of  non-interacting spins now present on the lattice. 
In addition to this,  the nearest-neighbor and next-nearest-neighbor
hopping terms must also both have pre-factors of $\delta$ to reflect
this finite probability that a site to  which one of the two
interacting spins may hop could be occupied by one  of the
non-interacting spins.  Given all of these necessary changes, the
Hamiltonian (without the additional $V_{0}$- and $U_{C}$-potential
terms) should now be expressed as
\begin{eqnarray}
{\cal H}_{tt'JJ_3}^{} & = & 
     -\delta t \sum_{{\bf r},\bbox{\delta},\sigma}
          c_{{\bf r},\sigma}^\dagger
          c_{{\bf r}+\bbox{\delta},\sigma}^{}
     -\delta t' \sum_{{\bf r},\bbox{\gamma},\sigma}
          c_{{\bf r},\sigma}^\dagger
          c_{{\bf r}+\bbox{\gamma},\sigma}^{}
          \label{eq:tt2JJ3coopham} \\
 & & -\frac{J}{4} \sum_{{\bf r},\bbox{\delta},\sigma}
          (n_{{\bf r},-\sigma}^{} n_{{\bf r}+\bbox{\delta},\sigma}^{}
          \nonumber \\  & & ~~~~~~~~~~
          +c_{{\bf r}+\bbox{\delta},\sigma}^\dagger
          c_{{\bf r},-\sigma}^\dagger
          c_{{\bf r}+\bbox{\delta},-\sigma}^{}
          c_{{\bf r},\sigma}^{}) \nonumber \\
 & & -\delta \frac{J_3}{4}
          \sum_{{\bf r},\bbox{\delta'}\neq\bbox{\delta},\sigma}
          (c_{{\bf r}+\bbox{\delta},\sigma}^\dagger
          n_{{\bf r},-\sigma}^{} c_{{\bf r}+\bbox{\delta'},\sigma}^{}
          \nonumber \\ & & ~~~~~~~~~~~~~~~~ 
          +c_{{\bf r}+\bbox{\delta},\sigma}^\dagger
           c_{{\bf r},-\sigma}^\dagger
           c_{{\bf r}+\bbox{\delta'},-\sigma}^{}
           c_{{\bf r},\sigma}^{}), \nonumber
\end{eqnarray}

Upon solving for $a({\bf r})$ and self-consistently  solving for
$a(0)$ and the four $a(\delta)$'s, two solutions are again obtained
of the forms
\begin{eqnarray}
a_s({\bf r}) & = & J \left( 1 + 3 \delta \frac{J_3}{J} \right)
  a(\bbox{\delta}_{1})
  \frac{1}{(2 \pi)^2} \int d{\bf k} e^{i{\bf k}\cdot{\bf r}}
   \nonumber \\ & & \times \frac{
  \left( \cos{k_{x}} + \cos{k_{y}} - 2 I_{1}/I_{0} \right)}
  {\delta \xi({\bf k}) - E/2 + U_C({\bf k})}, \\
a_{d}({\bf r}) & = & J \left( 1 - \delta \frac{J_3}{J} \right)
  a(\bbox{\delta}_{1})
  \frac{1}{(2 \pi)^2} \int d{\bf k} e^{i{\bf k}\cdot{\bf r}}
  \nonumber \\ & & \times \frac{
  \left( \cos{k_{x}} - \cos{k_{y}} \right)}
  {\delta \xi({\bf k}) - E/2 + U_C({\bf k})},
\end{eqnarray}
which once again correspond to extended $s$-wave and $d$-wave
solutions,  respectively..  Here, the limit $V_{0} \rightarrow \infty$
has already been done.  Now the definition of $U_C({\bf k})$ from
Eq.~(\ref{eq:Vk}) can be used,  and the limit of $U_C^0 \rightarrow
\infty$ is taken, yielding the final solutions
\begin{eqnarray}
a_s({\bf r}) & = & J \left( 1 + 3 \delta \frac{J_3}{J} \right)
  a(\bbox{\delta}_{1})
  \frac{1}{(2 \pi)^2} \int_{\xi({\bf k})>0} d{\bf k}
  e^{i{\bf k}\cdot{\bf r}}
  \nonumber \\ & & \times \frac{
  \left( \cos{k_{x}} + \cos{k_{y}} - 2 I_{1}/I_{0} \right)}
  {\delta \xi({\bf k}) - E/2},
\label{eq:asr4} \\
a_{d}({\bf r}) & = & J \left( 1 - \delta \frac{J_3}{J} \right)
  a(\bbox{\delta}_{1})
  \frac{1}{(2 \pi)^2} \int_{\xi({\bf k})>0} d{\bf k}
  e^{i{\bf k}\cdot{\bf r}}
  \nonumber \\ & & \times \frac{
  \left( \cos{k_{x}} - \cos{k_{y}} \right)}
  {\delta \xi({\bf k}) - E/2}.
\label{eq:adr4}
\end{eqnarray}
Integrations are, therefore, done over those states lying outside of
the  Fermi surface;  the filled states do not contribute to the
interactions of the problem.

The relationship for $J$ and $E$ can be found for the finite density
problem  as well as the expectation values of the various terms of the
extended  $t$-$J$ Hamiltonian.  The equations connecting $J$ and $E$
can be expressed as
\begin{equation}
\frac{1}{J} = B_{J_3} \frac{1}{(2 \pi)^2} \int_{\xi({\bf k})>0}
  d{\bf k} \cos{k_{x}} A_{\bf k},
\label{eq:JEcoop}
\end{equation}
where now
\begin{eqnarray}
  A_{\bf k} & = & (\cos{k_{x}} + \cos{k_{y}} - 2 I_{1}/I_{0})/
     (\delta \xi({\bf k}) - E/2), \nonumber \\
  B_{J_3} & = & 1 + 3 \delta J_3/J
\end{eqnarray}
for $s$-wave and
\begin{eqnarray}
  A_{\bf k} & = & (\cos{k_{x}} - \cos{k_{y}})/
     (\delta \xi({\bf k}) - E/2), \nonumber \\
  B_{J_3} & = & 1 - \delta J_3/J
\end{eqnarray}
for $d$-wave. The integrals in Eq.~(\ref{eq:JEcoop}) cannot be done
analytically. The actual  calculations entailed a simple Monte Carlo 
integration method where special care was taken when $E$ was  small
and the denominator of the integrand tended towards zero along the
Fermi surface.  

These eigenvalue-type equations can again be shown to  be equivalent
to the stability form of the gap equation at zero temperature.  The
difference with the dilute case is that the interaction
$V_{\text{gap}}$ which appears in the gap equation  now has a
frequency cutoff
\begin{equation}
V_{\text{gap}}({\bf k}, {\bf k'}) = \left\{
\begin{array}{ll}
V({\bf k}, {\bf k'}), & \xi({\bf k}) > 0 \\
0, & \xi({\bf k}) < 0.
\end{array} \right.
\end{equation}
This looks artificial.  However, the  Cooper bound state equation in
the conventional case has the same asymmetrical cutoff.  The
difference between the asymmetrical cutoff and the more symmetrical
cutoff of the BCS interaction is in fact responsible for the factor of
two difference in the exponential factors in the  Cooper binding
energy and the BCS gap.  Again, this difference does not influence the
{\it stability} issue of interest here.
 
The normalization is found to be
\begin{eqnarray}
\left< \psi | \psi \right> & = & \sum_{{\bf r}_1,{\bf r}_2}
    |a({\bf r}_1-{\bf r}_2)|^{2} \nonumber \\
    & = &
    J^{2} B_{J_3}^{2} |a(\bbox{\delta}_{1})|^2
    \frac{N_s}{(2 \pi)^{2}}
    \int_{\xi({\bf k})>0} d{\bf k} A_{k}^{2},
\end{eqnarray}
and the expectation values of the terms may be expressed
\begin{eqnarray}
\left<{\cal H}_{t}\right> & = & \frac{\left<\psi|{\cal H}_{t}|\psi\right>}
     {\left<\psi|\psi\right>} \nonumber \\
     & = &
     -4 \delta t \frac {\int_{\xi({\bf k})>0} d{\bf k}
     A_{\bf k}^{2} (\cos{k_{x}} + \cos{k_{y}})}
     {\int_{\xi({\bf k})>0} d{\bf k} A_{\bf k}^{2}}, \\
\left<{\cal H}_{t'}\right> & = & \frac{\left<\psi|{\cal H}_{t'}|\psi\right>}
     {\left<\psi|\psi\right>} \nonumber \\
     & = &
     -8 \delta t' \frac {\int_{\xi({\bf k})>0} d{\bf k}
     A_{\bf k}^{2} \cos{k_{x}} \cos{k_{y}}}
     {\int_{\xi({\bf k})>0} d{\bf k} A_{\bf k}^{2}}, \\
\left<{\cal H}_{J}\right> & = & \frac{\left<\psi|{\cal H}_{J}|\psi\right>}
     {\left<\psi|\psi\right>} \nonumber \\
     & = &
     -\frac{4}{B_{J_3}} \frac {\int_{\xi({\bf k})>0} d{\bf k}
     A_{\bf k} \cos{k_{x}}}
     {\int_{\xi({\bf k})>0} d{\bf k} A_{\bf k}^{2}}, \\
\left<{\cal H}_{J_3}\right> & = & \frac{\left<\psi|{\cal H}_{J_3}|\psi\right>}
     {\left<\psi|\psi\right>} \nonumber \\
     & = &
     -\frac{12}{B_{J_3}} \left(\delta \frac{J_3}{J} \right)
     \frac{\int_{\xi({\bf k})>0} d{\bf k} A_{\bf k} \cos{k_{x}}}
     {\int_{\xi({\bf k})>0} d{\bf k} A_{\bf k}^{2}}, \text{$s$},
     \\ & = &
     +\frac{4}{B_{J_3}} \left(\delta \frac{J_3}{J} \right)
     \frac{\int_{\xi({\bf k})>0} d{\bf k} A_{\bf k} \cos{k_{x}}}
     {\int_{\xi({\bf k})>0} d{\bf k} A_{\bf k}^{2}}, \text{$d$}.
\end{eqnarray}


\begin{figure}[ht]
\begin{center}
\leavevmode
\epsffile{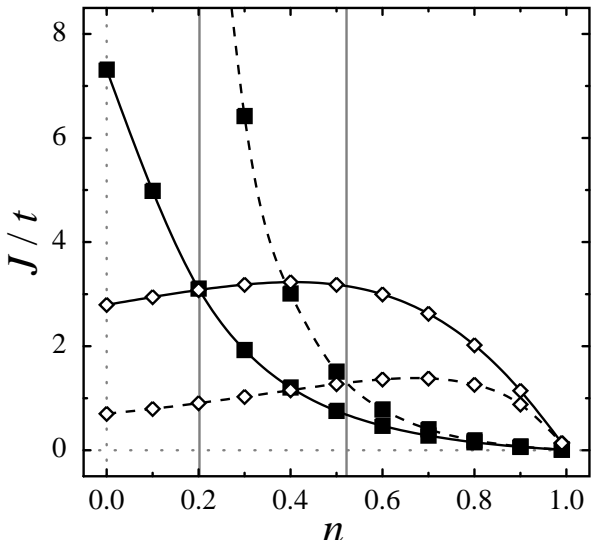}
\caption[]{
Phase diagram of coupling constant $J$ vs.\ spin density $n$ in the
Cooper  problem for binding energy $E = -0.01~t$, $J_3 = 0$ (solid 
line), and $J_3 =  J$ (dashed line).  The cross-over points between
$d$- (closed squares)
and $s$-wave (open diamonds) occur at specific  densities for both cases,
where these densities remain fairly fixed as $E$  (and accordingly $J$)
is decreased even more.  With the addition of the three-site term,
$s$-wave commands more of the phase diagram while $d$-wave is suppressed.}
\label{fig:phdiag2}
\end{center}
\end{figure}

Equation (\ref{eq:JEcoop}) is divergent in the limit $E \rightarrow
0$.  In  other words, given the presence of a Fermi surface ($\delta <
1$ and $n > 0$), $J$ does  not have a finite critical value as it did
in the dilute limit ($\delta =  1$ and $n = 0$).  This means that
there exists a bound state ($E < 0$) for any potential $J$ no matter
how small.  This is indeed the conclusion  of the Cooper problem:  any
pairing potential has a bound state for one  pair of spins, and this
must extend to the case of multiple pairs of  spins.  It is now
important to see which of the two symmetries has the  largest binding
energy $E$ (in absolute value) for a given $J$ as both $t'$ and
$\delta$ are varied.  In reality, $J$ is expressed as an integral 
containing $E$, and so $E$ is fixed as $J$ is calculated for varying 
values of both $t'$ and $\delta$.  Fig.~\ref{fig:phdiag2} shows the 
diagram of $J$ vs.~$n$ for a fixed binding energy $E = -0.01~t$ for
$t' = 0$ and $J_3 = 0$. The curves for $s$-wave and $d$-wave cross at
a  density of about $n = 0.2$.  This means that $d$-wave has the
larger  binding energy at smaller potential $J$ for densities $n >
0.2$, whereas $s$-wave has the larger binding energy at smaller $J$
for $n <  0.2$.  Turning this statement around, given a set $J$,
$d$-wave would have  the higher binding energy for $n > 0.2$ and
$s$-wave would be dominant for $n < 0.2$.  Indeed, if the  curves of
$J$ vs.~$n$ (as calculated from the above derivations) were to be
plotted as in Fig.~\ref{fig:phdiag2}  for decreasing values of the
binding energy ($E \rightarrow 0$), this cross-over point would remain
relatively fixed even as the  $J$-curves collapsed to $J = 0$ over all
$n$ at $E = 0$.  This cross-over between an $s$- and $d$-wave phase 
near $n = 0.2$ has been seen by Dagotto, {\it et al.}, in a quite 
different low-density study \cite{dagotto}.


\begin{figure}[t]
\begin{center}
\leavevmode
\epsffile{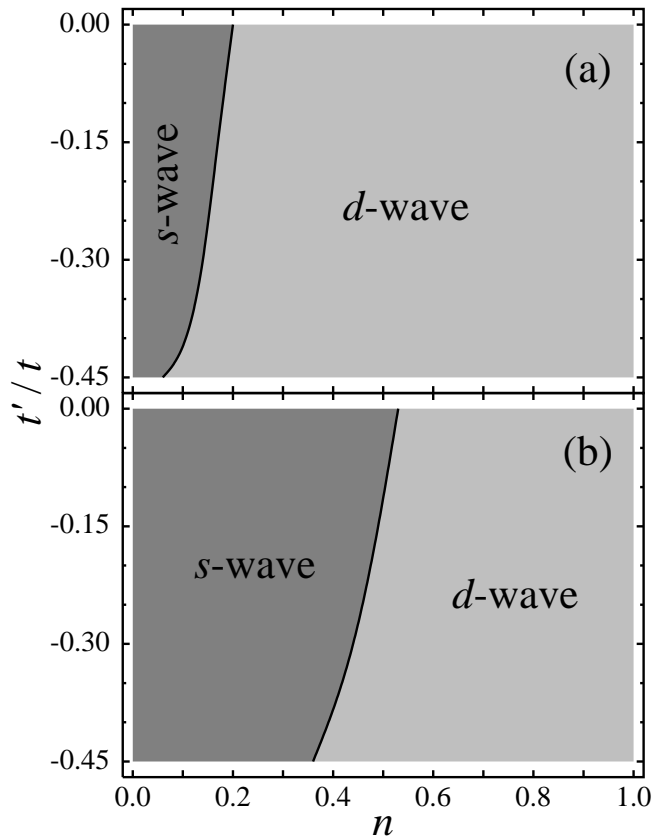}
\caption[]{
Phase diagram of next-nearest-neighbor hopping parameter $t'$ vs.\
spin  density $n$ in the Cooper problem for (a) $J_3 = 0$ and (b) $J_3
= J$.   The phase boundary represents small binding energy $E$ (and,
therefore,  small coupling constant $J$).  These diagrams show that
the $t'$-term  enhances $d$-wave over $s$-wave, whereas the $J_3$-term
enhances $s$-wave  over $d$-wave.}
\label{fig:phdiag3}
\end{center}
\end{figure}

The interesting trends of these calculations can be seen when the 
various parameters are changed.  When the three-site term is added, or 
$J_3 = J$, this cross-over point shifts to the right in the phase 
diagram as seen in Fig.~\ref{fig:phdiag2} to around $n = 0.53$.
Here,  $s$-wave commands more of the phase diagram while $d$-wave is 
suppressed.  This is consistent with the previous results of
Sec.~\ref{sec:dilute} where the  three-site term was found to be an
$s$-wave enhancing interaction.  The  cross-over points were also
found for various values of $t'$.  These  results are  shown in
Fig.~\ref{fig:phdiag3}.  Here, the cross-over from $d$-wave to 
$s$-wave with decreasing $n$ (increasing $\delta$) occurs at lower $n$
as  the $t'$-term increases in strength.  Once again it is seen that
the  next-nearest-neighbor hopping term enhances $d$-wave, where this
symmetry  takes over  more of the phase diagram as $t'$ increases in
magnitude.  It can also be  seen even more clearly in
Fig.~\ref{fig:phdiag3} that the addition of the  three-site term
increases the size of the $s$-wave phase as has now been  described in
great detail.

The solution of the Cooper problem shows that the presence of a Fermi
surface can stabilize $d$-wave over $s$-wave.  To see this in more
detail,  note that the interaction in momentum space is $V({\bf k},
{\bf k'}) = -(3/4) (\cos(k_x-k_x')+\cos(k_y-k_y'))$ and the  potential
energy is proportional to 
\begin{equation}
\sum_{{\bf k},{\bf k'}}V({\bf k},{\bf k'}) a({\bf k})a({\bf k'}).
\end{equation}
Now divide momentum-pair space into three regions: region (I), ${\bf
k} - {\bf k'} \approx 0$; region (II), ${\bf k} - {\bf k'} \approx
(\pm \pi, \pm \pi) $; region (III), ${\bf k} - {\bf k'} \approx (2
\pi, 0)$ or $(0,2 \pi)$. In region (I), $V({\bf k},{\bf k'}) > 0$, in
region (II), $V({\bf k},{\bf k'}) < 0$, and in region (III), $V({\bf
k},{\bf k'}) > 0$. Furthermore in region (I), $a_d({\bf k})a_d({\bf
k'}) \geq 0$, in region (II), $a_d({\bf k})a_d({\bf k'}) \leq 0$, and
in region (III), $a_d({\bf k})a_d({\bf k'}) \geq 0$, for the $d$-wave
case.  For the $s$-wave case, $a_s({\bf k})a_s({\bf k'}) \geq 0$ in
all regions. In general, when regions (I) and (III) are important,
then $s$-wave will have lower potential energy.  When region (II) is
important, then $d$-wave has lower potential energy.  The presence of
the Fermi surface enhances the stability of the $d$-wave solution
because it reduces the importance of region (I) and increases the
importance of region (II).

Although the dilute limit and the Cooper problem are not truly 
realistic pictures of superconductivity in the high-$T_{c}$ cuprates, 
these rather simplistic pictures do yield important behavior of the 
$t$-$J$ model which extends into the many-body problem of multiple
pairs  of interacting spins.  This crossing-over from  $d$-wave to
$s$-wave suggests regions of parameter space where $d$-wave  and
$s$-wave could even compete and form a kind of mixed $s$-$d$  state.
Section \ref{sec:meanfield} includes analysis of the mean field
$t$-$J$ model, which is shown indeed to support mixed $s$-$d$-pairing.


\section{The Mean Field Approach}
\label{sec:meanfield}

A next logical step in the analysis of the extended $t$-$J$ model  is
to evaluate its mean field Hamiltonian.  As with the dilute  limit and
the Cooper problem, the goal is to gain physical insight  into the
contributions of the various terms of the model, but now  by using a
more complex picture than just a simple single-pair  model.  Of most
interest is the form and symmetry of the gap  function $\Delta({\bf
k})$ as obtained from the zero-temperature BCS gap equation. To derive
$\Delta({\bf k})$, the full Hamiltonian Eq.~(\ref{eq:tt2JJ3ham}) must
first be reduced to the form
\begin{eqnarray}
  {\cal H}_{\text{mf}} & = &
    \sum_{{\bf k},\sigma} \varepsilon({\bf k}) n_{{\bf k},\sigma}
    \nonumber \\
  & &
    + \sum_{{\bf k},{\bf k'}} V({\bf k},{\bf k'})
    c_{{\bf k}\uparrow}^\dagger c_{-{\bf k}\downarrow}^\dagger
    c_{-{\bf k'}\downarrow}^{} c_{{\bf k'}\uparrow}^{}.
  \label{eq:mfham}
\end{eqnarray}
(Throughout this discussion, $J_3 = J$.) As with the Cooper problem,
factors of $\delta$ are included to account for the restriction of
no-double-occupancy since ${\cal H}_{\text{mf}}$ does not take this
explicitly into account. Equation (\ref{eq:mfham}), as applied to the
mean field $t$-$J$  model, is thus produced by transforming
Eq.~(\ref{eq:tt2JJ3coopham}) into a Hamiltonian of creation and
annihilation operators in  momentum space.  $\varepsilon({\bf k})$ is
once again the  tight-binding energy dispersion, and $V({\bf k},{\bf
k'})$ is of  the form
\begin{eqnarray}
  V({\bf k},{\bf k'}) & = &
    - J (\cos(k_x)\cos(k'_x) + \cos(k_y)\cos(k'_y)) \label{eq:Vkk} \\
  & &
    - \delta J_3 \left[ (\cos(k_x)\cos(k'_x) + \cos(k_y)\cos(k'_y))
    \right. \nonumber \\
  & & ~~~~
    + 2 \left. (\cos(k_x)\cos(k'_y) + \cos(k_y)\cos(k'_x)) \right] .
    \nonumber
\end{eqnarray}

With $V({\bf k},{\bf k'})$ now defined, ${\cal H}_{\text{mf}}$ can be
evaluated using the BCS wavefunction
\begin{equation}
  \left| \psi \right> = \prod_{\bf k} (u_{\bf k} + v_{\bf k}
  c^\dagger_{{\bf k}\uparrow} c^\dagger_{-{\bf k}\downarrow})
  \left| 0 \right>,  
\end{equation}
where the BCS variational parameters are of the usual forms
\begin{eqnarray}
  u_{\bf k}^2 & = & \frac{1}{2} \left( 1 -
    \frac{\xi({\bf k})}{\sqrt{\xi^2({\bf k}) + \Delta^2({\bf k})}}
    \right), \\
  v_{\bf k}^2 & = & \frac{1}{2} \left( 1 +
    \frac{\xi({\bf k})}{\sqrt{\xi^2({\bf k}) + \Delta^2({\bf k})}}
    \right).
\end{eqnarray}
Here, $\xi({\bf k}) = \varepsilon({\bf k}) - \mu$ and $\Delta({\bf
k})$ must be obtained from the zero-temperature gap  equation
Eq.~(\ref{eq:gapequ}).

The gap function $\Delta({\bf k})$ was calculated from the
self-consistent BCS gap  equation using an iterative numerical method. 
A matrix of initial  gap function values was defined on a grid
representing points in  momentum space.  A first iteration matrix was
then calculated from  the gap equation using the initial matrix as
input.  A linear  combination of the first iteration and initial
matrices was then  defined and optimized with regards to the gap
equation, and this  process was repeated until the matrices converged
upon a steady  solution.  The integrity of  the gap function solutions
was checked ({\it i}) by using a variety  of initial matrices,
including matrices of random-valued elements,  and ({\it ii}) by
changing the size of the matrices.  The  iterative method did indeed
converge on reproduceable, self-consistent solutions by using lattice
sizes of $442$ to  $1682$ $k$-states.

The gap function for the mean field $t$-$J$ model was found to always
have  three solutions which satisfied the self-consistent
zero-temperature BCS gap equation.  The first was the trivial solution
$\Delta({\bf k}) = 0$, and the other two were of the familiar forms
\begin{eqnarray}
  \Delta_s({\bf k}) & = & \psi_s (\cos k_x + \cos k_y), \\
  \Delta_d({\bf k}) & = & \psi_d (\cos k_x - \cos k_y),
\end{eqnarray}
where the magnitudes $\psi_s$ and $\psi_d$ vary as functions of $J$,
$\delta$ and $t'$. Under certain circumstances, there were also found
to be $s$+$d$ solutions when $\psi_s$ and $\psi_d$ were comparable in
size, when $J$ was larger than some  threshold value of approximately
$1.0~t$, and when $\delta$ was of  some intermediate value $0.15$ to
$0.45$.  Hence, in at least a  qualitative fashion, $sd$-mixing is
supported by  the mean field $t$-$J$ model in regions of parameter
space where  the $s$- and $d$-wave phases are competitive.


\begin{figure}[ht]
\begin{center}
\leavevmode
\epsffile{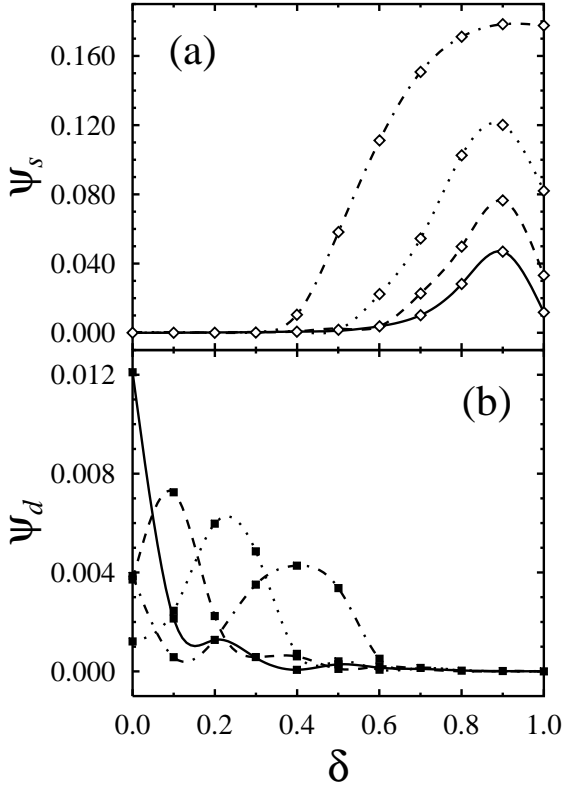}
\caption[]{
Magnitudes of gap equation solutions to the BCS gap equation vs 
hole density for $J_3 = J = 0.33~t$.  Shown are (a) $s_{x^2+y^2}$- and 
(b) $d_{x^2-y^2}$-wave gap amplitudes for $t'$ values of $0.00~t$
(solid lines), $-0.15~t$ (dashed lines), $-0.30~t$ 
(dotted lines), and $-0.45~t$ (dot-dashed lines).  The increase in 
$\psi_s$ in (a) with $t'$ is due to the flattening of the energy 
dispersion at low spin density (large hole density) as well as the 
narrowing of the band width as negative $t'$ increases in value.
The $\psi_d$ peaks in (b) lie at hole dopings which closely match
optimal doping in the materials modeled by the corresponding peaks'
$t'$ values.}
\label{fig:mags}
\end{center}
\end{figure}

More interesting results were obtained from the mean field gap 
calculations where physical values of the variational parameters  were
used.  For $J = 0.33~t$, there were no significant mixed state
solutions found for any values of $\delta$ or $t'$, but the $s$- and
$d$-wave solutions yielded surprising  trends.  Figure \ref{fig:mags}
shows the magnitudes of these  solutions to the mean field gap
equation over many hole densities  and values of $t'$.  The $s$-wave
magnitudes are largest around  lower spin density as expected from the
Cooper problem results, however the values increase with  more
negative $t'$.  This trend of $t'$ enhancing $s$-wave seems  contrary
to the previous results from the dilute limit and Cooper  problem, but
this incongruent result can be explained by the  nature of the energy dispersion. 
At these low spin densities, the  filled states lie close the the
energy minimum.  As $t'$ becomes  more negative, the energy dispersion
becomes more flat around the  minimum, and the band gap reduces in
width.  These two properties  of the energy dispersion become
overpowering effects on the $s$-wave gap function's magnitude for
larger values of $\delta$, despite  $t'$ becoming more negative.
This is an expected density-of-states effect, already pointed 
out by several authors \cite{dagotto,newns}.

The $d$-wave magnitudes of Fig.~\ref{fig:mags}(b) are the most 
striking results of all.  The magnitudes are highest closer to 
half-filling, again as expected from the Cooper problem results, 
where $\delta$ has values more representative of actual physical 
superconducting systems.  For a given value of $t'$, the highest gap
function magnitude seems to occur at a hole density representative of
optimal doping for the particular superconducting material being 
modeled by that $t'$ value.

For example, the highest gap function  magnitude for $t' = -0.15~t$
occurs at $\delta = 0.15$, which is  close to optimal doping for
La$_{2-x}$Sr$_x$CuO$_4$, whereas for  $t' = -0.45~t$, the highest
magnitude lies at $\delta = 0.4$,  close to optimal doping for
YBa$_2$Cu$_3$O$_{7-x}$.

Further mean field calculations were made with regard to 
anisotropy.  The previous results were all made with tetragonal 
symmetry, where the relative phase between the $s$- and $d$-wave 
components of the solutions to the gap equation was found to be 
$\pi/2$ and $0$ when the starting matrix contained complex and 
real random values, respectively.  The ground state energies for 
both cases were found to be degenerate, at least for the parameter 
ranges used in these calculations.  When an anisotropic 
energy dispersion was assumed, namely
$\epsilon({\bf k}) = - 2 t_a \cos{k_x} - 2 t_b \cos{k_y}
- 4 t' \cos{k_x} \cos{k_y}$,
where $t_a \neq t_b$, this relative 
phase between the $s$- and $d$-wave components was only found to be 
$0$, even when an initial matrix of random complex values was 
used.  As the anisotropy increased, $\psi_s$ increased and $\psi_d$ 
decreased, however, both effects were not profound.

A systematic study of the extended $t$-$J$ model has been presented in
the last three sections, beginning with a simplistic single-pair model
in the dilute limit, followed by evaluation of a single-pair in the
Cooper problem, and then ending with a mean field approach.  At each
step along this evolution, we have shown that $s$- and $d$-wave phases
do exist (and indeed co-exist in a mixed fashion), dependent upon the
shape and size of the Fermi surface as described by the parameters
$t'$ and $\delta$. Variational Monte Carlo calculations have already
shown the  presence of a mixed $s$-$d$ symmetry in the $t$-$J$ model
for the case of  $t' = 0$ \cite{koltenbah}.  A cross-over was seen
where the ground state  symmetry was pure $d$-wave for $\delta < 0.12$
and an  admixture of $s$- and $d$-wave for $\delta > 0.12$ up to
around $\delta =  0.4$. Further unpublished calculations show that the
addition of the  $t'$-term decreases and eliminates this mixed phase
as $t'$ increases in  magnitude.  The trends derived from the dilute
limit, the Cooper  problem, and the mean field approach as treated in
the $t$-$J$ model have now yielded physical reasons for the results
seen in the VMC calculations.

Support for  the applicability of the extended $t$-$J$ model in
describing the  high-temperature superconductors must far more
importantly come from  experimental evidence for a gap symmetry
dependent upon Fermi surface  shape and doping.  A survey of various
experimental work is made in  Sec.~\ref{sec:survey} where the case is
made for identifying  the extended $t$-$J$ model as a detailed
material-specific description of  superconductivity in the cuprates.


\section{Experimental Survey}
\label{sec:survey}

The calculations reported in this paper identify two trends for the 
${\it ground~state}$ of high-$T_c$ superconductors. There is
dependence of the form of the gap function on the dispersion relation
quantified by the second-neighbor hopping. There is dependence on the
hole concentration.  These are parameters which vary considerably from
system to system. The theory therefore indicates that not all
high-$T_c$ superconductors are alike.  Experiments on magnetic
properties also indicate differences between
the different systems;  the incommensurability of the
short-range order in the La$_{2-x}$Sr$_x$CuO$_4$
system does not appear to be present in the 
YBa$_2$Cu$_3$O$_7$ system.  Similarly, these properties are doping-dependent;
the various signatures of the spin gap are present in 
underdoped, but not optimally doped YBa$_2$Cu$_3$O$_{7-\delta}$.
(For a review of magnetic properties, see
Ref.~ \cite{arno}.)  This section is devoted to a
survey of different systems to see if there is experimental support
for this point of view.
Indeed, a growth of theoretical 
work has resulted from these recent experimental results, the details of 
which we present in Sec.~\ref{sec:survey}.  According to the above results, 
we should investigate each separate
high-$T_c$ for $d$-wave, $s$-wave, and $s$-$d$ mixed wave
\cite{gabi,gua}. The last possibility requires further discussion.
$s$-$d$ mixing can occur in two forms: $s$+$id$ and $s$+$d$, depending
upon whether the proportionality constant between them does or does
not have an imaginary part, respectively. The calculations above do
not discriminate between these two possibilities.  For recent work on
the mixing hypothesis in various models, see Refs.~\cite{carbotte,karen}. 
Variational Monte Carlo calculations, which could  in
principle show an energy difference in the ground state as a function
of the phase difference, were inconclusive \cite{koltenbah}.   In the
$s$+$id$ state there are no nodes in the gap, and time reversal
symmetry is broken.  Josephson interference experiments put very
strong upper limits on the admixture of $s$-wave in any  $s$+$id$ state
\cite{vh}. In addition, the $s$+$id$ state would lead to domain
effects which are  not observed.  (By contrast,  the negative results
in optical dichroism experiments do {\it not} test for the $s$+$id$
state \cite{qpl}.)

In the following discussion, however, we shall concentrate on the
$s$+$d$ state as the most likely  mixed pairing state, and all 
references to mixed-wave pairing below are to this state only.  In this state
the $d$-wave gap nodes, located along the $k_x = \pm k_y$ directions,
move toward the crystal axis $k_x= 0$ or $k_y = 0$ as the relative
weight of the $s$-component increases. Which of the two directions is
chosen depends on the sign of the coupling to the orthorhombic
distortion which exists in most high-$T_c$ materials. If there is no
such distortion, then $s+d$ and $s-d$ are degenerate.  If a critical
size of the $s$-wave component is reached, then the nodes move
together and they annihilate one another.  Beyond that point, the
system has a ``hard'' gap; there are no  quasiparticle excitations. 
If the $s$-wave component predominates, the $d$-wave admixture may be
thought of as contributing some orthorhombic distortion to a gap with
otherwise tetragonal symmetry.  Further discussion of the mixed state
is found in Ref.~\cite{koltenbah}.

The purpose of this section is to survey what is known about the gap
symmetry of {\it different} high-$T_c$ materials, without making the
assumption that the gap symmetry must be the same in all.   Previous
surveys have often lumped all materials together. This would defeat
our purpose of looking for material-specific trends. A more complete
survey, but from a different point of view, may again be found in
Ref.~\cite{doug}.

The best-studied compound is of course YBa$_2$Cu$_3$O$_7$. It appears
from temperature-dependent penetration depth measurements  $\lambda
(T)$ to be quite clear that there are low energy quasiparticle
excitations in this system \cite{hardy}. This shows that any admixture
of $s$-wave must be less than the critical value. Anisotropy in
$\lambda$ suggests that  pure $d$-wave is not the gap symmetry in this
material, but this could also be due simply to the 
orthorhombic character of the crystal structure.

The Josephson interference experiments of the Illinois and ETH groups
measure the difference in the phase of electrons passing through
perpendicular faces of a crystal \cite{illinois,eth}.  It is rather
accurately measured to be $\pi$, and this is in agreement with
independent  measurements.  This is likely to  reflect very well the
phase shift under a $90^{\circ}$ rotation in momentum space.  It is
important to point out that this is consistent with pure $d$-wave {\it
or} with the $s$+$d$ state, both of which give $\pi$ for this
quantity.   These experiments rule out the $s$+$id$ state,  which has
a shift of $\pi/2$ if the two components  have equal weight \cite{vh}. 
The IBM tricrystal experiments \cite{ibm} further show that the nodes
in the gap cannot be more than $30^{\circ}$ 
from the $k_x =\pm k_y$ points. This puts 
constraints on the weight of any admixture of $s$-wave. 
$c$-axis tunneling experiments, on the other hand, would suggest that
the $s$-wave component is not negligible \cite{dynes}.  Indeed, if the
critical current  along this direction is related to the size of the
$s$-wave gap,  the $s$-wave part is perhaps 10\% to 20\% of the
$d$-wave part. Nuclear magnetic resonance experiments on this material
are inconsistent with a pure $s$-wave state, and have often been
interpreted in terms of $d$-wave superconductivity  \cite{bulut}. The
$s$+$d$ interpretation has not been done; however, the fact that
anisotropic $s$-wave \cite{sudip} and $s$+$id$ \cite{koltenbah}
theories can  account for the data equally well suggests strongly 
that the observations are consistent with $s$+$d$ as well. 
Temperature-dependent Raman scattering on YBa$_2$Cu$_3$O$_7$
shows most sensitivity in the $B_{1g}$ geometry \cite{hackl}.
This supports $d$-wave pairing \cite{mz,tpd1}.
Some limitation can be placed on the proportion of 
s-wave from these experiments, but how much is not clear.  
Quasiparticle (Giaever)
tunneling experiments performed by point contacts
\cite{ybctun}, tend to show the presence of states near the chemical
potential, consistent with $d$-wave or $s$+$d$, but it has generally
proven to be difficult to unambiguously compare theory and experiment
(and even experiment and experiment) in this area.  It appears that 
quasiparticle tunneling performed in conjunction with 
scanning tunneling microscopy holds out the best hope of
gap determination, as surface quality and site specificity
are important issues \cite{kk}.  
At present, however, this method gives ambiguous results for the
density of states in the superconducting state of 
YBa$_2$Cu$_3$O$_7$ \cite{nantoh}. 


\begin{table}
\twocolumn[\hsize\textwidth\columnwidth\hsize\csname %
@twocolumnfalse\endcsname
\centering
\caption{\bf EXPERIMENTAL SURVEY}
\label{tab:survey}
\begin{minipage}{4.0in}
\begin{tabular}{||c||c|c|c|c|c||}
              &     Y123      & B22128 & B2212(8+$x$) & LSCO & NDCO \\
\hline \hline
$\lambda(T)$  & $d$, $d$+$s$  &   -    &      -       &  -   & $s$  \\
\hline
Jos.~Interf.  & $d$, $d$+$s$  &   -    &      -       &  -   &  -   \\
\hline
Photemiss.    &      -        &  $d$   &   $d$+$s$    &  -   &  -   \\
\hline
NMR           & $d$, $d$+$s$  &   -    &      -       &  -   &  -   \\
\hline
Tunneling     &    $d$ ?      & $d$ ?  &      -       &  -   & $s$  \\
\hline
Neutr.~Scatt. &      -        &   -    &      -       & $s$  &  -   \\
\hline
$t'/t$        &    -0.4       & -0.2 ? &    -0.2      &-0.16 &+0.16
\end{tabular}
\end{minipage}
]
\end{table}

Using 
Ginzburg-Landau theory, Berlinsky and Kallin studied the Abrikosov vortex 
lattice and found that the lattice changes from a triangular to an 
oblique to a square spacing with increasing field and $sd$-mixing 
parameter \cite{berlinsky}.  
This may explain neutron scattering experiments in YBa$_2$Cu$_3$O$_7$
\cite{keimer}.  The interesting point about this work
from the point of view of this paper
is that the $s$-wave component of the order parameter is
zero in the absence of current flow.  Instead, it is induced by 
the supercurrent which is itself a response to the applied field \cite{rjhc2}.
Thus, there does not need
to be an actual $s$-wave instability in the 
(translationally invariant) ground state for the effects of
$s$-$d$ mixing to be experimentally observable.  It may be sufficient
that there exists a phase boundary between pure $d$ wave and 
and $s$-$d$ mixed phase and that some systems are close to it.

The second material in terms of 
detailed gap studies is Bi$_2$Sr$_2$CaCu$_2$O$_{8+x}$.
Its particular interest is that the gap
is visible in angle-resolved photoemission.  
Although this experiment 
measures only the absolute magnitude of the gap function, it has an
angular resolution not available in the Josephson tunneling
experiments. Furthermore, $x$ can be 
varied, and this is of pivotal importance from our point of view, as
the prediction that the $s$-wave component can develop as the
hole doping ($x$) increases can be tested.  The {\it optimally doped}
compound shows a well-developed gap $\Delta_x$ along at least one of
the crystal axes (say, $k_x=0$), and a very small or zero gap
$\Delta_{xy}$ along the diagonal (say, $k_x = \pm k_y$) \cite{shen}.
This is in contrast to results in {\it overdoped}
Bi$_2$Sr$_2$CaCu$_2$O$_{8+x}$ ($T_c = 75 K$) \cite{ma}.  Here it is
found that there is a gap along the $k_x = \pm k_y$ direction at lower
temperatures, and that the anisotropy ratio $\Delta_{x}/\Delta_{xy}$
is therefore temperature-dependent. This has been successfully
analyzed under the hypothesis of $s$+$d$ and is inconsistent with pure
$d$-wave \cite{joebob}.  Raman spectroscopy in this compound
also supports d-wave pairing, again with no clear limitation on
$s$-$d$ mixing \cite{tpd2}.
Giaever tunneling experiments are not always consistent with one another in
Bi$_2$Sr$_2$CaCu$_2$O$_{8+x}$ materials,  but again often suggest states
near the chemical potential.  Recent scanning tunneling
microscopy experiments have been seen as indicative of an $s$+$d$
state \cite{ichimura}, a $d$ state \cite{renner}, and dependence on
surface layer is also observed \cite{mura}.

Experiments probing gap structure are somewhat harder to come by in
other materials.  Nd$_{2-x}$Ce$_x$CuO$_4$ has been the subject of
experiments measuring the temperature dependence  of the penetration
depth \cite{maryland} and the tunneling spectra \cite{johnz}.  Both
are consistent with a fully-developed gap of $s$-wave type. Neutron
scattering experiments in La$_{2-x}$Sr$_x$CuO$_4$ detect a gap in the
$k_x =  k_y$ direction \cite{mason}.  This experiment is consistent
with $s$-wave or $s$+$d$, though the somewhat isotropic
suppression of scattering would favor pure $s$-wave, probably of
a gapless nature.  Impurity scattering in a $d$-wave superconductor
can also produce rather isotropic suppression \cite{qs}.
The tricrystal experiment has recently been performed on
a Tl-based film \cite{tsuei}.  It also indicates that nodes in the 
gap of this material must be within $30^{\circ}$ of the diagonals in the
Brillouin zone.

We summarize this information in Table \ref{tab:survey}. Also given in
the table are values of $t'$, the second neighbor hopping element,
taken from various sources \cite{tohyama,fehske,dessau}. The theory
would generally predict that $s$-wave should  be more likely as one
moves to the right in the table, as $t'$ decreases in magnitude.  The doping is not
so easy to estimate, but it is certain that the overdoped
Bi$_2$Sr$_2$CaCu$_2$O$_{8+x}$ has more holes than the optimally doped
Bi$_2$Sr$_2$CaCu$_2$O$_{8+x}$, and therefore should lie in the table
as shown.  In general, the trend is indeed toward more $s$-wave
behavior on the right hand side of the table, in qualitative agreement
with theory.

A common experimental objection to the $s$-$d$ mixed state is that
it involves two superconducting transitions, but only one is observed.
This subject is treated rather thoroughly in Ref.~\cite{koltenbah}
and we summarize their conclusions here.  Symmetry analysis shows that there
are two transitions only if the material is tetragonal.  In the
orthorhombic case, $s$-wave and $d$-wave have the same transformation
properties under the operations of the point group and only one
transition will occur.  In UPt$_3$, the only material in 
which two superconducting transitions are known to
occur as a function of temperature,
they are visible in specific heat measurements and in 
ultrasonic absorption and velocity 
measurements.  In high-$T_c$ systems, the 
transition in specific heat has been clearly seen only 
in YBa$_2$Cu$_3$O$_7$, which is quite orthorhombic.
In the nearly-tetragonal Bi-based and Tl-based systems,
the appropriate measurements have not been done.


\section{Conclusion}
\label{sec:conclusion}

The calculations presented in this paper use the simple mean-field
method for the investigation of superconductivity.  There is, in
practice, little choice of methods if the Hamiltonian has a form which
is flexible enough to show material-specific behavior. (See, for
example, Eq.~(\ref{eq:tt2JJ3coopham}).)  More sophisticated methods,
such as slave-boson methods with gauge fluctuations, have been
developed for the investigation of $t$-$J$-type models, but these are
quite difficult to handle even for the simplest models. Finite-size
calculations are limited to lattices whose linear dimensions are
comparable to, or smaller than, a coherence length. Quantum Monte
Carlo (QMC) calculations  (again on simpler models than the current
one) have severe sign problems in the regimes of interest.  VMC
calculations may be possible, but this is left for future
investigation. To the extent that we have been able to compare our
results with the  VMC results, there is qualitative agreement.  
Finally, our calculations in the spin-interaction model are cruder
than  the strong-coupling calculations which have been done in the
spin-fluctuation model.  The spin-interaction model is genuinely
microscopic, however, not semiphenomenological. Compensating for  the
disadvantage of the approximate treatment of interactions is the
advantage that the method is physically transparent - the reasons for
the trends which appear are all evident, and much of the paper has
been devoted to the explanation of these trends.  This transparency is
not always  a feature of  highly numerical methods such as QMC or VMC. 

Thus the trends towards $d$-wave with increasing $t'$ can be seen as
softening angular  variations in the pair wavefunction, whereas the
trend towards $s$-wave with increasing doping is  seen as due to a
relaxation of phase-space restrictions.  The trend towards
higher optimal doping with increasing $t'$ is a
density-of-stated effect.  There is evidence for all of 
these effects in experiments.  The most dramatic of these is the change from
strong $d$-wave in YBa$_2$Cu$_3$O$_7$ to $s$-wave in Nd$_{2-x}$Ce$_x$CuO$_4$. 

This analysis allows us to evaluate qualitatively  the actual
relevance of the calculations  to real materials.  For example, let
the ideal Fermi surface  of the Cooper problem be replaced by the
actual momentum distribution.  The latter will be much flatter, with a
relatively small discontinuity at the Fermi surface (assuming a Fermi
liquid normal state) or no discontinuity (Luttinger liquid normal
state). Therefore the phase space restrictions will be weaker in the
real system. This would favor $s$-wave and move the crossing point of
$s$-wave and $d$-wave at  $n = 0.55$ (at $t' = 0$) in
Fig.~\ref{fig:phdiag2} to higher values with increasing $t'$.
Relaxation of the very large $U$  (strong {\it short range} Coulomb
interaction) approximation will also favor $s$-wave, as the constraint
of  absolutely no double-occupancy has no effect on $d$-wave, whereas
it suppresses $s$-wave pairing. By the same token, the neglect of the
long range Coulomb interaction in our model probably prejudices our
results toward $d$-wave.  The usual frequency-mismatch arguments so
important for superconductivity with electron-phonon interaction do
not apply here, since all interactions are instantaneous. It is
precisely for that reason that our results fly in the face of
conventional wisdom that magnetic mechanisms can lead only to $d$-wave
pairing.  In the absence of a cutoff in the interaction, all of
momentum space comes into play, and the kinetic energy of the pair
plays an important role. This brings $s$-wave superconductivity into
the picture.

The trends which arise from our simple analysis of this flexible model
appear to describe some of the otherwise puzzling and contradictory
results on different high-$T_c$ materials.  They show that the ground
state of different materials may be different in the context of a
magnetic mechanism.

We would like to thank T.~M.~Rice, S.~Hellberg, A. Sudb\o, M. Ma, P.W.
Anderson and M. Randeria for useful discussions and correspondence.
This work was supported by the  National Science Foundation through
Grant DMR-9214739 and by NORDITA, Copenhagen, Denmark


\end{document}